\documentclass[journal]{IEEEtran}
\usepackage{amsmath,amsfonts,amssymb}
\usepackage{algorithmicx}
\usepackage[ruled]{algorithm2e}
\usepackage{array}
\usepackage[caption=false,font=normalsize,labelfont=sf,textfont=sf]{subfig}
\usepackage{textcomp}
\usepackage{stfloats}
\usepackage{url}
\usepackage{verbatim}
\usepackage{graphicx}
\usepackage{cite}
\usepackage[table]{xcolor}
\usepackage{booktabs}
\usepackage{pifont}
\newcommand{\cmark}{\ding{51}}
\usepackage{multirow}
\usepackage{makecell}
\usepackage{threeparttable}
\usepackage{caption}
\usepackage{bm}
\usepackage{float}
\usepackage[hidelinks]{hyperref}
\newcommand{\MethodName}{CGSA}
\newsavebox{\hyperfigbox}

\begin{document}
\title{Fault Diagnosis of Dynamic Systems Under Unknown Operating Conditions: \\A Condition-Guided Selective Adaptation Approach}
\author{Jiaming Liu, Zeyi Liu, Hongshuo Zhao, Pengyu Han, Xiao He,~\IEEEmembership{Senior Member,~IEEE}
\thanks{This work was supported in part by National Natural Science Foundation of China under grants 62525308, 624B2087, 62473223, and 52172323, in part by Beijing Natural Science Foundation under grant L241016 (\emph{Corresponding author: Xiao He.})}
\thanks{Jiaming~Liu is with the School of Computer Science and Engineering, Beihang University, Beijing 100191, P. R. China (email: 23371007@buaa.edu.cn).}
\thanks{Zeyi~Liu, Pengyu~Han, and Xiao~He are with the Department of Automation, and the Institute for Embodied Intelligence and Robotics, Tsinghua University, Beijing 100084, China (emails: liuzy21@mails.tsinghua.edu.cn, hpy24@mails.tsinghua.edu.cn, hexiao@tsinghua.edu.cn).}
\thanks{Hongshuo~Zhao is with the MCC5 Group Shanghai Co. LTD 201900, ShangHai, China (email: zhaohongshuo@sh5mcc.com).}
}

\markboth{}{}
\maketitle

\begin{abstract}
	Fault diagnosis under unknown operating conditions remains challenging for dynamic industrial systems, 
	as the distribution shift caused by changing operating conditions can significantly degrade the performance of diagnostic models in real-world applications.
	To address the problem, a condition-guided selective adaptation approach is proposed.
	Specifically, a novel continuous operating-condition adversarial learning strategy with progressive training is developed in the offline stage to enhance the generalization ability of the diagnostic model. 
	During online deployment, residual operating-condition responses are exploited to identify reliable unlabeled samples from streaming data, which are then used to update the diagnostic model.
	Extensive experiments on real-world gearbox and motor datasets have demonstrated that the proposed framework outperforms state-of-the-art methods in diagnostic accuracy while maintaining relatively low test-time, 
	showing its potential for practical industrial applications.
\end{abstract}


\section{Introduction}
\IEEEPARstart{W}{ith} 
the advancement of modern technology, industrial systems are becoming increasingly complex and dynamic, 
while industrial equipment typically operates under constantly changing conditions, such as load and speed~\cite{Lzy-IL-Review,Complexity-2,Lzy-AL,wang2024fault}. 
Such dynamic systems introduce complex nonlinear and non-stationary characteristics into signals \cite{Non-stationary-2,Nonlinear}, 
increasing the difficulty of diagnosis \cite{dong2023neural,FD-Meaning-2}.
In this context, 
\textit{multi-condition fault diagnosis} (MCFD) has become a promising direction for facilitating the transition of fault diagnosis methods from controlled laboratory environments to practical deployment \cite{HpyMCFD},
with significant implications for maintaining system stability and reducing economic losses \cite{ma2023adaptive,FD-Meaning-4,zhong2023overview,FD-Meaning-3}.
\color{black}

Specifically, traditional machine learning models usually rely on the \textit{independent and identically distributed} (i.i.d.) assumption \cite{Not-i.i.d-2, Data-driven-3,Zhang2020}.
However, changes in operating conditions can cause a substantial distribution shift between training and deployment data, 
which significantly reduces the performance of diagnostic models in practical applications \cite{Not-i.i.d-3,Not-i.i.d-4}.
Meanwhile, operating conditions in dynamic systems are continuously evolving and difficult to exhaustively cover through data collection in advance \cite{Hard2Collect-1, Hard2Collect-2, Hard2Collect-3, LI2025130137}. 
As a result, sufficient samples are usually unavailable for previously unknown operating conditions \cite{Insufficient-Samples-2, Insufficient-Samples-3}. 
To address these issues, MCFD aims to enable diagnostic models to maintain reliable performance when the data distribution changes, 
or even when the target operating conditions are previously unknown.

Among existing MCFD methods, \textit{domain adaptation} (DA) and \textit{domain generalization} (DG) are two representative paradigms.
DA methods are designed to mitigate distribution shifts by aligning features between the labeled source domain and the unlabeled target domain \cite{DA-based-2}. 
\color{black}
DG methods go a step further by extracting domain-invariant features to improve generalization ability using only source-domain data, 
thus eliminating the dependence on target domain data \cite{DG-based-4}.
However, most existing MCFD methods focus primarily on offline knowledge transfer, and their models remain fixed after deployment \cite{Offline-1,Offline-2,Offline-3,Song2024},
which means that model performance mainly relies on prior data acquired during the training phase.
\color{black}
Nevertheless, in real-world industrial scenarios, a large amount of streaming data under previously unknown operating conditions can be continuously collected during deployment.
Although these online data usually lack labels, they still provide a valuable opportunity for the model to perceive target-distribution changes and perform adaptive updates.
Therefore, the introduction of an online adaptation mechanism serves as an important way to enable the diagnostic model to dynamically adjust itself using continuously arriving streaming data
and ultimately enhance its robustness under unknown conditions \cite{Lzy-RTFD,Hpy-MPOS-RVFL}. 

In this context, \textit{online test-time adaptation} (OTTA) provides a promising paradigm \cite{OTTA-Survey-1},
which focuses on dynamically updating a deployed model using only unlabeled online data streams without access to source-domain data \cite{OTTA-Survey-2}. 
Along this emerging direction, several recent studies have attempted to introduce OTTA into fault diagnosis scenarios \cite{OTTA-instance-1,OTTA-instance-2,OTTA-instance-3}.
The problem is that OTTA methods were originally proposed for test-time distribution shifts in computer vision tasks, such as style or weather changes,
where category semantics and domain variation factors are generally considered independent.
In contrast, operating conditions in industrial systems are closely tied to the physical dynamics of the monitored equipment. 
Fault occurrence may perturb vibration patterns and transmission relationships, 
causing the actual operating conditions to ultimately deviate from the preset operating conditions \cite{Fault-Condition-Coupling-1,Fault-Condition-Coupling-2,Fault-Condition-Coupling-3}. 
As a result, fault-related and condition-related information can be naturally coupled in the feature space.
Furthermore, online data streams in industrial scenarios often exhibit distinct stage-wise structures, 
where the equipment usually operates in the healthy state before fault occurrence. 
Once a specific fault occurs, subsequent samples tend to remain within the same fault type until maintenance or shutdown, 
rather than alternating among healthy and different fault states.
Nevertheless, current OTTA-based fault diagnosis methods still rarely consider these characteristics explicitly.
Simply enforcing condition-invariant representations may distort fault-discriminative features \cite{Han2025RethinkingTR}. 
Moreover, without mechanisms tailored to such stage-wise streams, the diagnostic model may gradually overfit to the currently dominant sample category, eventually leading to error accumulation and catastrophic forgetting.
Therefore, it is necessary to develop an OTTA-based method that takes full account of the characteristics of industrial systems in order to better solve the problem of fault diagnosis under unknown operating conditions.

In this paper, a \textit{Condition-Guided Selective Adaptation} (\MethodName) framework is proposed to address the challenge of fault diagnosis under unknown operating conditions.
{\MethodName} explicitly considers the coupling between fault states and operating conditions in dynamic industrial systems. 
Specifically, a novel continuous operating-condition adversarial learning strategy with progressive training is developed in the offline stage to utilize labeled data from continuous operating conditions,
thereby enhancing cross-condition generalization without overly suppressing fault-discriminative information. 
Because the inherent fault-condition coupling makes strict condition invariance difficult to achieve,
the residual condition information is preserved, which is further considered as a reliability indicator for online sample selection.
During deployment, the source pre-trained model is frozen to leverage coupling information to guide the student-model adaptation. 
The main contributions can be summarized as follows:
\begin{enumerate}

\item A continuous operating-condition adversarial learning strategy with progressive training is developed for the offline stage. 
By exploiting labeled data collected under continuous operating conditions, 
the cross-condition generalization of the model is enhanced.

\item An online adaptation mechanism is designed to provide reliable updates and mitigate error accumulation.
The source pre-trained model generates residual condition responses and pseudo-labels for identifying reliable samples,
which are then used to update the student model.

\item Extensive experiments on real-world gearbox and motor datasets demonstrate that the proposed approach achieves superior diagnostic performance
and desirable test-time efficiency, showing its potential for practical industrial applications.

\end{enumerate}

\section{The Proposed Method}
\subsection{Problem Formulation}
For a dynamic system with constantly changing operating conditions, 
a fault diagnosis task with \(K\) diagnostic classes is considered, including the healthy state and multiple fault states.
In the offline stage, a labeled training dataset 
$\mathcal{T}^{(0)}=\left\lbrace \bm{X}^{(0)}, \bm{y}^{(0)}, \bm{c}^{(0)}\right\rbrace$ can be collected under continuously changing operating conditions, 
where $\bm{X}^{(0)} \in \mathbb{R}^{n_0 \times d}$ contains $n_0$ samples and $d$-dimensional features, 
$\bm{y}^{(0)} \in \{1,2,\dots,K\}^{n_0}$ represents the corresponding $K$ diagnostic classes,
$\bm{c}^{(0)} \in \mathbb{R}^{n_0}$ indicates the continuous operating condition index.
In the online stage, neither class labels nor operating-condition information is available.
\color{black}
The model can only access an unlabeled data stream 
$\mathcal{S}=\{\bm{X}^{(1)},\bm{X}^{(2)},\ldots,\bm{X}^{(n)}\}$, 
where each batch $\bm{X}^{(t)} \in \mathbb{R}^{N \times d}$ contains $N$ samples.

The objective of this work is to establish a robust fault diagnosis framework for dynamic systems under unknown operating conditions.
The proposed method comprises two stages: an offline training stage and an online test-time adaptation stage.
\color{black}
In the offline stage, the model exploits continuous operating-condition information to enhance its generalization ability.
During the online test-time adaptation stage, the proposed method aims to update the diagnostic model effectively and timely, 
so as to adaptively improve its diagnostic performance under unknown operating conditions and meet the immediacy requirements of real-world scenarios. 

\begin{figure*}[!t]
	\centering
	\includegraphics[width=6.9in, keepaspectratio]{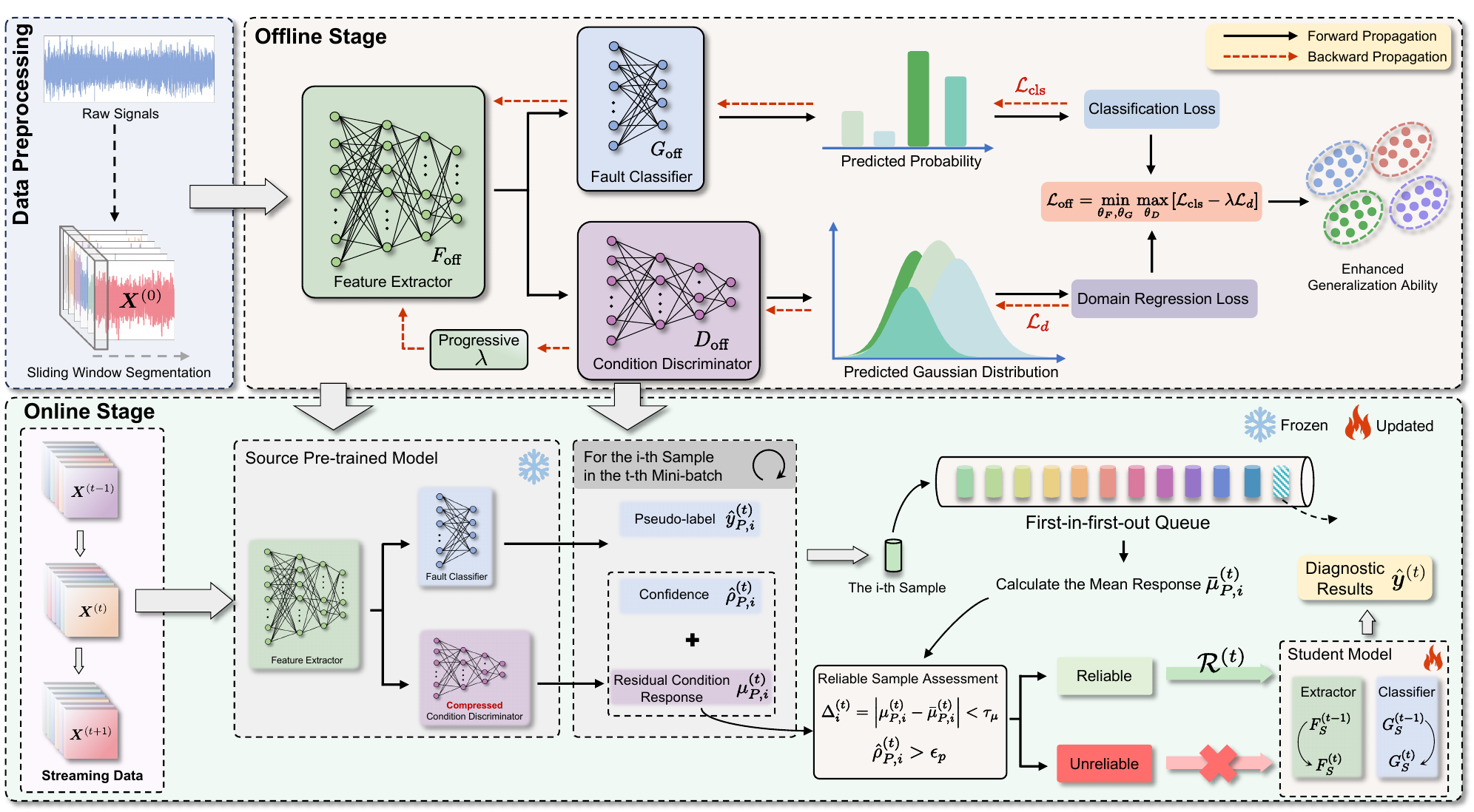}
	\caption{The overall flowchart of the proposed {\MethodName} approach.}
	\label{flowchart}
\end{figure*}

\subsection{Offline Stage}
In the offline stage, a continuous operating-condition adversarial learning strategy with progressive training, 
inspired by continuously indexed domain adaptation \cite{pmlr-v119-wang20h}, 
is adopted to improve the generalization ability of the diagnostic model.

Let $F_{\mathrm{off}}$, $G_{\mathrm{off}}$, and $D_{\mathrm{off}}$ denote the feature extractor, the fault classifier, and the condition discriminator, respectively. 
Different from conventional domain-adversarial learning methods that treat domain prediction as a discrete classification task, 
the condition discriminator is formulated as a probabilistic regressor with a dual-branch architecture.
Specifically, the mean branch predicts the mean $\mu_d$, 
while the log-variance branch predicts the log-variance $\log \sigma_d^2$ of a Gaussian distribution from the extracted feature representation $F_{\mathrm{off}}(\bm{x})$.
These two parameters are then used to characterize the continuous operating-condition information.

The domain regression loss is defined based on the Gaussian negative log-likelihood, which encourages the discriminator to better fit the true continuous condition value by modeling its predictive distribution:
\begin{equation}
    \mathcal{L}_{d}
    =
    \mathbb{E}_{(x,c)\sim\mathcal{T}^{(0)}}
    \left[
    \frac{1}{2}\log\sigma_d^2
    +
    \frac{(c-\mu_d)^2}{2\sigma_d^2}
    \right].
\end{equation}
Meanwhile, the classification objective is formulated as:
\begin{equation}
    \mathcal{L}_{\mathrm{cls}}=
	\mathbb{E}_{(x,y)\sim \mathcal{T}^{(0)}}
	\left[
	\mathcal{L}_{\mathrm{ce}}
	\left(
	G_{\mathrm{off}}(F_{\mathrm{off}}(x)), y
	\right)
	\right],
\end{equation}
where $\mathcal{L}_{\mathrm{ce}}(\cdot)$ denotes the cross-entropy loss.

During adversarial optimization, the condition discriminator is optimized to
infer the continuous operating condition index from the learned 
representations, whereas the feature extractor is trained to confuse the 
condition discriminator. Accordingly, the feature extractor is encouraged 
to learn condition-insensitive representations while preserving 
fault-discriminative information. The offline adversarial optimization is 
formulated as:
\begin{equation}\label{eq_loss_off}
    \mathcal{L}_{\mathrm{off}}
    =
    \min_{\theta_F,\theta_G}
    \max_{\theta_D}
    \left[
    \mathcal{L}_{\mathrm{cls}}
    -
    \lambda \mathcal{L}_{d}
    \right],
\end{equation}
where $\theta_F$, $\theta_G$, and $\theta_D$ denote the parameters of the feature extractor, the fault classifier, and the condition discriminator, respectively.

Considering the inherent coupling between fault-related and condition-related information, 
directly applying adversarial training may distort fault-discriminative features.
Therefore, a progressive adversarial coefficient is adopted to gradually increase the adversarial intensity, 
thereby improving the stability of training.
The coefficient is formulated as:
\begin{equation}
    \lambda=
    \frac{2}{1+\exp(-\gamma p)}-1,
\end{equation}
where $p$ denotes the training progress, and $\gamma$ controls the increasing rate.

Overall, the offline training stage improves the cross-condition generalization ability of the diagnostic model and provides a foundational model for online adaptation.

\subsection{Online Stage}
Although offline training strategies improve model performance,
offline domain generalization cannot fully guarantee consistently reliable diagnostic performance under all unknown operating conditions, especially when severe distribution shifts occur.
Therefore, an online test-time adaptation mechanism is introduced to further improve the adaptability of the diagnostic model during deployment.

For the initial online setup, the source pre-trained model remains fixed throughout the online stage and is not involved in parameter updates, 
while an adaptive student model is initialized from it for online update.
In detail, the fixed source pre-trained model consists of \(F_P\), \(G_P\), and \(D_P\), initialized from \(F_{\mathrm{off}}\), \(G_{\mathrm{off}}\), and the mean branch of \(D_{\mathrm{off}}\), respectively.
Meanwhile, the student model consists of \(F_S\) and \(G_S\), 
which are initialized from \(F_{\mathrm{off}}\) and \(G_{\mathrm{off}}\), respectively.

At time step \(t\), the student model first generates diagnostic predictions for the current mini-batch \(\bm{X}^{(t)}\) before model adaptation to meet real-time requirements. 
The predicted probability matrix is denoted by \(\mathcal{G}_{S}^{(t)}=G_{S}^{(t-1)}(F_{S}^{(t-1)}(\bm{X}^{(t)}))\), 
where \(\mathcal{G}_{S}^{(t)}\in\mathbb{R}^{B\times K}\). The diagnostic outputs of the current mini-batch are obtained as:
\begin{equation}
	\hat{\bm{y}}^{(t)}
	=
	\mathop{\arg\max}_{k\in\{1,\ldots,K\}}
	\mathcal{G}_{S}^{(t)}[:,k].
	\label{eq_output}
\end{equation}
\color{black}
Subsequently, the pre-trained model is used to provide stable reference predictions. 
For each sample \(\bm{x}_i^{(t)} \in \bm{X}^{(t)}\), the predicted probability vector of the pre-trained model is denoted as
\(\mathcal{G}_{P,i}^{(t)}=G_P(F_P(\bm{x}_i^{(t)}))\), 
where \(\mathcal{G}_{P,i}^{(t)}\in\mathbb{R}^{1\times K}\).
Based on the prediction vector, the pseudo-label \(\hat{y}_{P,i}^{(t)}\) 
and its corresponding confidence \(\hat{\rho}_{P,i}^{(t)}\) are obtained as:
\begin{equation}
	\hat{y}_{P,i}^{(t)}
    =
    \mathop{\mathrm{argmax}}\limits_{j\in\{1,\ldots,K\}}
    \mathcal{G}_{P,i}^{(t)}(j),
    \label{eq_pre-trained_y_t}
\end{equation}
and
\begin{equation}
	\hat{\rho}_{P,i}^{(t)}
    =
    \mathop{\mathrm{max}}\limits_{j\in\{1,\ldots,K\}}
    \mathcal{G}_{P,i}^{(t)}(j).
    \label{eq_pre-trained_rho_t}
\end{equation}

Furthermore, the pre-trained condition discriminator produces a residual condition response for the same sample,
which can be written as $\mu_{P,i}^{(t)} = D_P(F_P(\bm{x}_i^{(t)}))$.
It should be noted that $\mu_{P,i}^{(t)}$ should not be interpreted as a direct measurement of the physical operating condition.
Instead, it is regarded as a compressed residual response that retains and reflects the inherent fault-condition coupling information after offline training.
In fact, different residual condition response values can be associated with different specific faults.

In order to support the online update mechanism, a dynamic queue is maintained to record the status of recent samples. 
Before processing the \(i\)-th sample \(\bm{x}_{i}^{(t)}\) in the \(t\)-th mini-batch $\bm{X}^{(t)}$ in sequence, 
the queue is denoted by \(\mathcal{Q}_{i}^{(t)}\), which contains at most the \(W\) most recent residual condition responses generated by the pre-trained discriminator. 
At the beginning of an online stream, the queue is initialized as empty, 
and the residual responses of the first \(W\) samples are sequentially inserted to warm up the queue without sample evaluation or model updating.
Once the queue reaches its capacity, the reliability of each subsequent sample is evaluated using the responses stored in the queue.
After evaluation, \(\mu_{P,i}^{(t)}\) is inserted into the queue in a first-in-first-out (FIFO) manner.

The mean response of the current queue is used to characterize the recent residual condition information:
\begin{equation}\label{eq_queue_mean}
    \bar{\mu}_{P,i}^{(t)}
    =
    \frac{1}{W}
    \sum_{\mu \in \mathcal{Q}_{i}^{(t)}} \mu,
\end{equation}
where \(\bar{\mu}_{P,i}^{(t)}\) is the mean of the latest \(W\) residual responses preceding \(\bm{x}_{i}^{(t)}\).
\color{black}

To evaluate the reliability of each incoming sample, the deviation between its residual condition response and the mean response stored in the queue is calculated.
Although $\mu_{P,i}^{(t)}$ is not a direct measurement of the physical operating condition, it preserves mild residual information induced by the inherent fault-condition coupling. 
Accordingly, for a locally continuous stream, a sample with a correct pseudo-label is generally more consistent with its corresponding fault-related feature region and is therefore more likely to produce a residual response compatible with this recent trajectory. 
In contrast, a misclassified sample may disrupt this local consistency because its features tend to fall into a feature-space regions associated with the incorrectly predicted fault classes, finally resulting in a larger residual deviation from $\bar{\mu}_{P,i}^{(t)}$.
The queue mean serves as a local temporal reference rather than a direct measure of sample similarity. 
A sample with a sufficiently small deviation is considered consistent with the recent prediction trajectory and is therefore regarded as reliable for online adaptation.
\color{black}
Based on the above criterion, the reliability deviation of the $i$-th sample in the $t$-th mini-batch is defined as:
\begin{equation}\label{eq_reliability_deviation}
    \Delta_i^{(t)}
    =
    \left|
    \mu_{P,i}^{(t)}
    -
    \bar{\mu}_{P,i}^{(t)}
	\right| .
\end{equation}
A smaller $\Delta_i^{(t)}$ indicates that the sample is more consistent with the recent prediction results.

To further improve the quality of selected samples, a confidence filtering criterion is introduced. 
By combining residual condition consistency with pseudo-label confidence filtering, the reliable sample set at time step $t$ is constructed as:
\begin{equation}\label{eq_reliable_set}
    \mathcal{R}^{(t)}
    =
    \left\{
    \bm{x}_{i}^{(t)}
    \mid
    \Delta_i^{(t)} < \tau_{\mu},
    \;
    \hat{\rho}_{P,i}^{(t)} > \epsilon_p
    \right\},
\end{equation}
where $\tau_{\mu}$ denotes the threshold for residual condition consistency, and $\epsilon_p$ denotes the pseudo-label confidence threshold.
Only samples satisfying both criteria are regarded as reliable and used for online adaptation, while the others are rejected to avoid generating unreliable gradients.

All reliable samples in the current mini-batch are used to update the student model. 
For each selected sample $\bm{x}_i^{(t)} \in \mathcal{R}^{(t)}$, the pseudo-label $\hat{y}_{P,i}^{(t)}$ is used as the supervision signal. 
The online loss is defined as:
\begin{equation}\label{eq_loss_on}
    \mathcal{L}_{\mathrm{on}}^{(t)}
    =
    \frac{1}{|\mathcal{R}^{(t)}|}
    \sum_{\bm{x}_i^{(t)} \in \mathcal{R}^{(t)}}
    \mathcal{L}_{\mathrm{ce}}
    \left(
    G_S^{(t-1)}
    \Bigl[
    F_S^{(t-1)}(\bm{x}_i^{(t)})
    \Bigr],
    \hat{y}_{P,i}^{(t)}
    \right).
\end{equation}
The student model is updated by online gradient descent:
\begin{equation}\label{eq_optimization}
\left\{
\begin{aligned}
    \theta_{F_S}^{(t)}
    &=
    \theta_{F_S}^{(t-1)}
    -
    \eta
    \nabla_{\theta_{F_S}}
    \mathcal{L}_{\mathrm{on}}^{(t)}, \\
    \theta_{G_S}^{(t)}
    &=
    \theta_{G_S}^{(t-1)}
    -
    \eta
    \nabla_{\theta_{G_S}}
    \mathcal{L}_{\mathrm{on}}^{(t)} ,
\end{aligned}
\right.
\end{equation}
where \(\eta\) denotes the learning rate for online adaptation, and
\(\theta_{F_S}\) and \(\theta_{G_S}\) denote the parameters of the student feature extractor
and student fault classifier, respectively.
If no reliable sample is selected from the current mini-batch, i.e.,
\(|\mathcal{R}^{(t)}|=0\), the student model is not updated.

\begin{algorithm}[t]
	\caption{Main Procedure of the Proposed Method.}
	\label{proposed_method}
	
	\KwIn{Offline dataset $\mathcal{T}^{(0)}=\{\bm{X}^{(0)},\bm{y}^{(0)},\bm{c}^{(0)}\}$; online stream $\mathcal{S}=\{\bm{X}^{(t)}\}_{t=1}^{\infty}$.}
	\KwOut{Prediction results $\hat{\bm{y}}^{(t)}$ at time $t$.}
	
	\textbf{Offline initialization:}
	Train $F_{\mathrm{off}}$, $G_{\mathrm{off}}$, and $D_{\mathrm{off}}$ with the offline objective in Eq.~\eqref{eq_loss_off}\;
	
	Keep the source pre-trained model fixed and initialize an adaptive student model from it;
	
	\textbf{Online stage:}
	\For{$t=1,2,\ldots$}{
		Receive an unlabeled mini-batch $\bm{X}^{(t)}$\;
		
		Initialize the reliable sample set $\mathcal{R}^{(t)}=\emptyset$\;

		Output the prediction $\hat{\bm{y}}^{(t)}$ using \eqref{eq_output}\;
		
		\ForEach{$\bm{x}_{i}^{(t)} \in \bm{X}^{(t)}$}{
			Obtain the pseudo-label $\hat{y}_{P,i}^{(t)}$ and confidence $\hat{\rho}_{P,i}^{(t)}$ using \eqref{eq_pre-trained_y_t} and \eqref{eq_pre-trained_rho_t}\;
			
			Obtain the residual condition response $\mu_{P,i}^{(t)}$;

			\If{$|\mathcal{Q}_{i}^{(t)}|=W$}{
    			Compute $\bar{\mu}_{P,i}^{(t)}$ via \eqref{eq_queue_mean}\;

    			Compute $\Delta_i^{(t)}$ via \eqref{eq_reliability_deviation}\;

				\If{$\Delta_i^{(t)} < \tau_{\mu}$ \textbf{and} $\hat{\rho}_{P,i}^{(t)} > \epsilon_p$}{
					Add $\bm{x}_{i}^{(t)}$ to $\mathcal{R}^{(t)}$\;
				}
			}

			Update the queue in a FIFO manner;
		}

		\If{$|\mathcal{R}^{(t)}|>0$}{
    	Compute the online loss $\mathcal{L}_{\mathrm{on}}^{(t)}$ via \eqref{eq_loss_on}\;
    
    	Update $F_S^{(t)}$ and $G_S^{(t)}$ using \eqref{eq_optimization}\;
		}
	}
\end{algorithm}

Following this process, the model can be continuously updated. 
The overall procedure of \MethodName\ is illustrated in Fig.~\ref{flowchart}, and the detailed steps are summarized in Algorithm~\ref{proposed_method}.

\section{Experiments}
\label{sec:experiments}
\subsection{General Experimental Settings}
\subsubsection{Datasets}
In this study, two self-collected datasets are employed to demonstrate the effectiveness of the proposed method in real-world dynamic systems, 
including the MCC5-THU Gearbox Dataset \cite{chen2024multi} and the MCC5-THU Motor Dataset \cite{CHEN2026112583}. 
Each dataset contains one healthy category and three fault categories,
and the collected data cover diverse operating conditions to enable a comprehensive evaluation.

\begin{figure}[!t]
	\centering
	\includegraphics[width=\columnwidth,keepaspectratio]{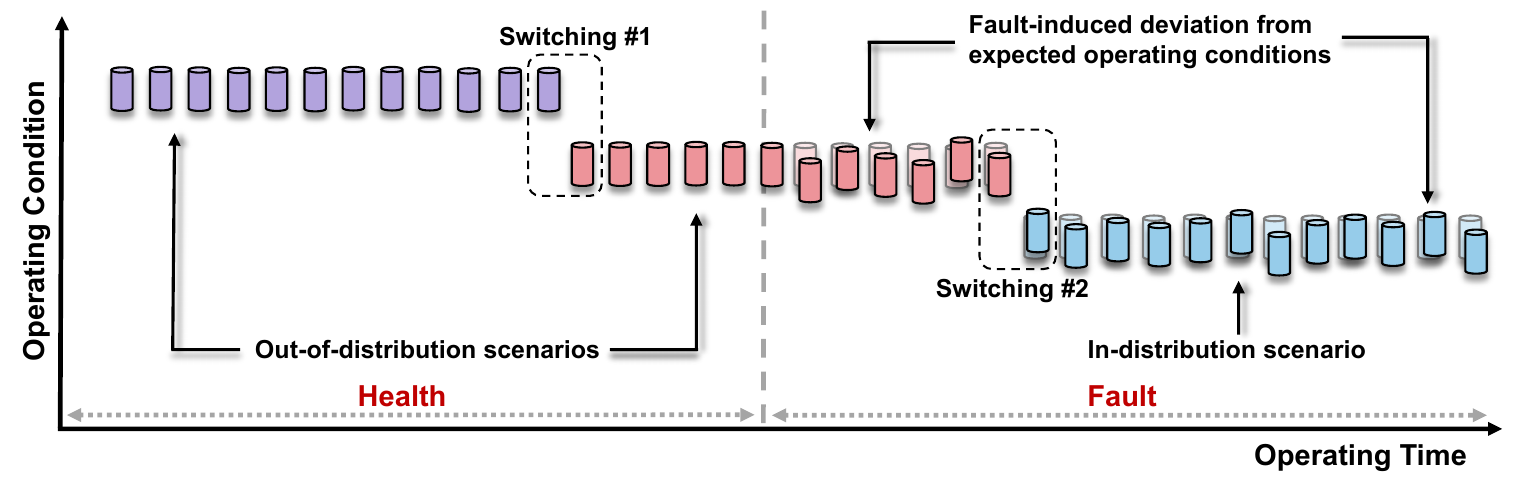}
	\caption{Illustration of the online data stream.}
	\label{fig:dataflow}
\end{figure}

\subsubsection{Evaluation Metrics}
For metrics, in addition to average accuracy, the Expected Calibration Error (ECE) is adopted to evaluate the reliability of model predictions. 
ECE measures the discrepancy between prediction confidence and actual accuracy, reflecting the calibration quality of the model \cite{EATA-C}.
A lower ECE value indicates better calibration performance.
Specifically, the prediction confidence interval $[0,1]$ is divided into $M$ equal bins, 
and the ECE is calculated as
\begin{equation}
    \mathrm{ECE}
    =
    \sum_{m=1}^{M}
    \frac{|B_m|}{N}
    \left|
    \mathrm{acc}(B_m)
    -
    \mathrm{conf}(B_m)
    \right|,
    \label{eq:ece}
\end{equation}
where $N$ denotes the total number of samples, 
$B_m$ denotes the set of samples whose prediction confidence falls into the $m$-th bin, 
and $|B_m|$ is the number of samples in $B_m$. 
The terms $\mathrm{acc}(B_m)$ and $\mathrm{conf}(B_m)$ represent the average accuracy and average confidence of samples in $B_m$, respectively. 
In this study, $M$ is set to 10.

\subsubsection{Implementation Details} 
\begin{table}[!htbp]
    \begin{center}
        \centering
        \caption{Construction of the offline dataset.}
        \label{tab:offline_condition_setting}
        \renewcommand{\arraystretch}{1.15}
        \setlength{\tabcolsep}{4.5pt}

        \begin{tabular}{c c c c c}
            \specialrule{0.1em}{1pt}{1pt}
            \specialrule{0.1em}{1pt}{2pt}
            \cmidrulewidth=0.5pt

            Dataset
            & Type
            & Segment $\mathcal{T}_1$
            & Segment $\mathcal{T}_2$
            & Segment $\mathcal{T}_3$ \\

            \specialrule{0.10em}{2pt}{2pt}

            \multirow{2}{*}{Gearbox}
            & Torque
            & $0\text{--}10$ Nm
            & $10$ Nm
            & $10\text{--}15$ Nm \\

            \cmidrule(lr){2-5}

            & Speed
            & \multicolumn{3}{c}{$1000$ rpm} \\

            \cmidrule[0.8pt](lr){1-5}

            \multirow{2}{*}{Motor}
            & Torque
            & $0\text{--}20$ Nm
            & $20$ Nm
            & $20\text{--}30$ Nm \\

            \cmidrule(lr){2-5}

            & Speed
            & \multicolumn{3}{c}{$2000$ rpm} \\

            \specialrule{0.1em}{2pt}{1pt}
            \specialrule{0.1em}{1pt}{1pt}
        \end{tabular}
    \end{center}
\end{table}
For the offline setup, both datasets are collected under fixed-speed settings, 
and their continuous condition indices are derived from the measured torque across three concatenated segments, as summarized in Table~\ref{tab:offline_condition_setting}. 
For each sample, the continuous condition label is derived from the corresponding measured torque and normalized as follows:
\begin{equation}
    c_i =
    \frac{T_i-T_{\min}}
         {T_{\max}-T_{\min}},
\end{equation}
where $T_i$ denotes the measured torque of the $i$-th sample,
and $T_{\min}$ and $T_{\max}$ are the minimum and maximum torque values used for normalization, respectively.
\begin{figure}[!t]
	\centering
	\subfloat[MCC5-THU gearbox test rig~\cite{chen2024multi}.]{%
		\includegraphics[width=0.7\linewidth,keepaspectratio]{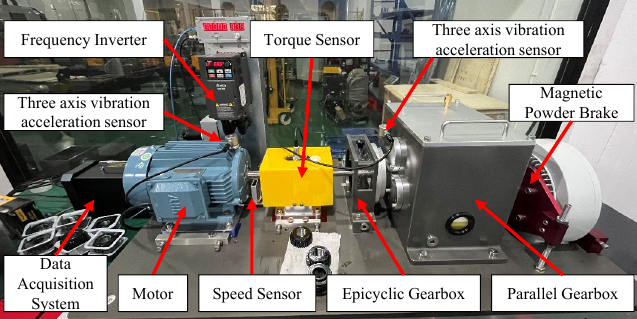}%
		\label{fig:mcc5-thu-gearbox}}\\
	\subfloat[MCC5-THU motor test rig~\cite{CHEN2026112583}.]{%
		\includegraphics[width=0.7\linewidth,keepaspectratio]{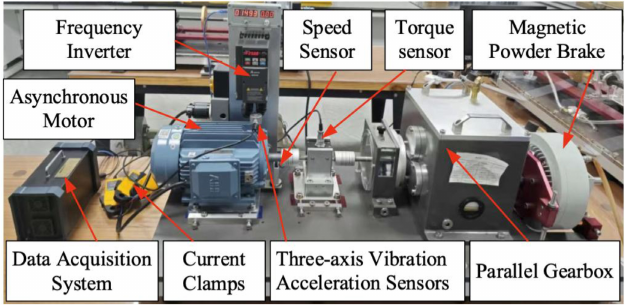}%
		\label{fig:mcc5-thu-motor}}
	\caption{Experimental test rigs.}
\end{figure}

To present a realistic operating scenario, each online data stream switches among three steady-state operating conditions, as reported in Table~\ref{tab:online_setting},
while the speed remains constant at the value used in the offline stage.
\color{black}
The first two conditions ($\mathcal{C}_1$ and $\mathcal{C}_2$) are treated as out-of-distribution (OOD) scenarios, 
while the last condition ($\mathcal{C}_3$) corresponds to an in-distribution (ID) scenario.
The ID scenario provides an important test stage for online methods, since continuous updates may lead to catastrophic forgetting.
In each online experiment, the data stream begins with healthy data, 
after which a specific fault is introduced at the midpoint of the stream, as illustrated in Fig.~\ref{fig:dataflow}.

\begin{table}[!htbp]
	\begin{center}
	  \centering
	  \caption{Details of online operating conditions.}
	  \label{tab:online_setting}
	  \renewcommand{\arraystretch}{1.15}
	  \setlength{\tabcolsep}{4.5pt}
	  \begin{tabular}{c c c c}
		  \specialrule{0.1em}{1pt}{1pt}
		  \specialrule{0.1em}{1pt}{2pt}
		  \cmidrulewidth=0.5pt
  
		  \multirow{2}{*}{Dataset}
		  & \multicolumn{2}{c}{OOD} & ID \\
		  \cmidrule(lr){2-3} \cmidrule(lr){4-4}
		   & Condition $\mathcal{C}_1$ & Condition $\mathcal{C}_2$ & Condition $\mathcal{C}_3$ \\
  
		  \specialrule{0.10em}{2pt}{2pt}
  
		  Gearbox
			& $20$ Nm   & $15$ Nm   & $5$ Nm \\
  
		  \cmidrule(lr){1-4}
  
		  Motor
			& $30$ Nm   & $40$ Nm   & $15$ Nm \\
  
		  \specialrule{0.1em}{2pt}{1pt}
		  \specialrule{0.1em}{1pt}{1pt}
	  \end{tabular}
	\end{center}
\end{table}
\color{black}

\subsubsection{Comparison Methods}
The 6-channel signals are first divided into mutually exclusive offline training and online test sets at the raw-data level to prevent data leakage.
Sliding-window segmentation is then independently performed within each set using a window size of 1024 and a step size of 16.
\color{black}
For DANN \cite{DANN}, the original three operating domains are directly used as discrete domain labels.
For the proposed method, the torque signal is normalized and used as the continuous operating condition label during offline adversarial training.

All methods are evaluated over 10 independent random trials using the same data partitions described in Section.~\ref{sec:experiments}, with the random seed initialized once to 10 for reproducibility. 
All experiments are conducted on a workstation equipped with 25 Intel Xeon Platinum 8470Q vCPUs and an NVIDIA GeForce RTX 5090 GPU.
\color{black}

\begin{table*}[!t]
	\begin{center}
	  \centering
	  \caption{The comparison results on gearbox Dataset regarding average Acc.($\uparrow$) and ECE($\downarrow$).}\label{table_comparison_gearbox}
	  \begin{threeparttable}
	  \footnotesize
	  \setlength{\tabcolsep}{1.0mm}
	  \resizebox{\textwidth}{!}{%
	  \begin{tabular}{c|c|c|ccccccccc}
		  \specialrule{0.1em}{1pt}{1pt}
		  \specialrule{0.1em}{1pt}{2pt}
		  \cmidrulewidth = 0.5em
		  \multirow{2.5}{*}{Fault Type} & \multirow{2.5}{*}{Stage} & \multirow{2.5}{*}{Metric} & \multicolumn{9}{c}{Methods} \\
		  \cmidrule(lr){4-12}
		  &  &  & \multicolumn{1}{c}{\textit{SupCon}} & \multicolumn{1}{c}{\textit{Tent-Full}} & \multicolumn{1}{c}{\textit{AdaContrast}} & \multicolumn{1}{c}{\textit{CoTTA}} & \multicolumn{1}{c}{\textit{DPLOT}} & \multicolumn{1}{c}{\textit{SURGEON}} & \multicolumn{1}{c}{\textit{OAFD}} & \multicolumn{1}{c}{\textit{Tri-Training}} & \multicolumn{1}{c}{\textit{CGSA}\textsuperscript{*}} \\
		\specialrule{0.10em}{2pt}{2pt}
		  \renewcommand{\arraystretch}{0.9}
		  \multirow{8}{*}{Gear Wear} & \multirow{2}{*}{$\mathcal{C}_1$} & Acc. & 0.735 $\pm$ 0.069 & 0.752 $\pm$ 0.070 & 0.800 $\pm$ 0.093 & 0.801 $\pm$ 0.077 & 0.791 $\pm$ 0.077 & 0.743 $\pm$ 0.070 & 0.746 $\pm$ 0.070 & 0.882 $\pm$ 0.040 & \mbox{\colorbox{gray!35}{0.891 $\pm$ 0.082}} \\[-0.28ex]
		   &  & ECE & 0.145 $\pm$ 0.019 & 0.143 $\pm$ 0.021 & 0.191 $\pm$ 0.024 & 0.185 $\pm$ 0.031 & 0.266 $\pm$ 0.055 & 0.145 $\pm$ 0.019 & 0.147 $\pm$ 0.021 & 0.207 $\pm$ 0.020 & \mbox{\colorbox{gray!35}{0.075 $\pm$ 0.070}} \\[-0.28ex]
		  \cmidrule(lr){2-12}
		   & \multirow{2}{*}{$\mathcal{C}_2$} & Acc. & 0.700 $\pm$ 0.036 & 0.668 $\pm$ 0.037 & 0.728 $\pm$ 0.054 & 0.687 $\pm$ 0.032 & 0.691 $\pm$ 0.034 & 0.686 $\pm$ 0.037 & 0.688 $\pm$ 0.039 & 0.738 $\pm$ 0.018 & \mbox{\colorbox{gray!35}{0.889 $\pm$ 0.028}} \\[-0.28ex]
		   &  & ECE & 0.053 $\pm$ 0.023 & 0.088 $\pm$ 0.034 & 0.087 $\pm$ 0.029 & 0.056 $\pm$ 0.028 & \mbox{\colorbox{gray!35}{0.040 $\pm$ 0.012}} & 0.068 $\pm$ 0.028 & 0.060 $\pm$ 0.024 & 0.053 $\pm$ 0.013 & 0.046 $\pm$ 0.019 \\[-0.28ex]
		  \cmidrule(lr){2-12}
		   & \multirow{2}{*}{$\mathcal{C}_3$} & Acc. & 0.996 $\pm$ 0.003 & 0.995 $\pm$ 0.004 & 0.995 $\pm$ 0.005 & 0.993 $\pm$ 0.005 & 0.996 $\pm$ 0.003 & 0.996 $\pm$ 0.004 & 0.995 $\pm$ 0.004 & \mbox{\colorbox{gray!35}{0.999 $\pm$ 0.000}} & 0.997 $\pm$ 0.004 \\[-0.28ex]
		   &  & ECE & 0.008 $\pm$ 0.004 & 0.007 $\pm$ 0.004 & 0.017 $\pm$ 0.011 & 0.020 $\pm$ 0.013 & 0.009 $\pm$ 0.005 & 0.007 $\pm$ 0.004 & 0.009 $\pm$ 0.005 & 0.006 $\pm$ 0.001 & \mbox{\colorbox{gray!35}{0.002 $\pm$ 0.002}} \\[-0.28ex]
		  \cmidrule(lr){2-12}
		   & \multirow{2}{*}{\textbf{Average}} & Acc. & 0.810 $\pm$ 0.029 & 0.805 $\pm$ 0.030 & 0.841 $\pm$ 0.038 & 0.827 $\pm$ 0.021 & 0.826 $\pm$ 0.030 & 0.808 $\pm$ 0.030 & 0.810 $\pm$ 0.030 & 0.873 $\pm$ 0.014 & \mbox{\colorbox{gray!35}{0.926 $\pm$ 0.035}} \\[-0.28ex]
		   &  & ECE & 0.048 $\pm$ 0.013 & 0.043 $\pm$ 0.018 & 0.080 $\pm$ 0.015 & 0.061 $\pm$ 0.019 & 0.094 $\pm$ 0.021 & 0.045 $\pm$ 0.015 & 0.044 $\pm$ 0.013 & 0.075 $\pm$ 0.008 & \mbox{\colorbox{gray!35}{0.040 $\pm$ 0.024}} \\[-0.28ex]
			 \cmidrule[0.8pt](r){1-12}
		  \multirow{8}{*}{Teeth Break} & \multirow{2}{*}{$\mathcal{C}_1$} & Acc. & 0.735 $\pm$ 0.069 & 0.752 $\pm$ 0.070 & 0.800 $\pm$ 0.093 & 0.795 $\pm$ 0.070 & 0.791 $\pm$ 0.077 & 0.743 $\pm$ 0.070 & 0.746 $\pm$ 0.070 & 0.882 $\pm$ 0.040 & \mbox{\colorbox{gray!35}{0.891 $\pm$ 0.082}} \\[-0.28ex]
		   &  & ECE & 0.145 $\pm$ 0.019 & 0.143 $\pm$ 0.021 & 0.191 $\pm$ 0.024 & 0.185 $\pm$ 0.031 & 0.266 $\pm$ 0.055 & 0.145 $\pm$ 0.019 & 0.147 $\pm$ 0.021 & 0.207 $\pm$ 0.020 & \mbox{\colorbox{gray!35}{0.075 $\pm$ 0.070}} \\[-0.28ex]
		  \cmidrule(lr){2-12}
		   & \multirow{2}{*}{$\mathcal{C}_2$} & Acc. & 0.747 $\pm$ 0.035 & 0.754 $\pm$ 0.043 & 0.708 $\pm$ 0.050 & 0.702 $\pm$ 0.046 & 0.735 $\pm$ 0.037 & 0.754 $\pm$ 0.039 & 0.742 $\pm$ 0.036 & 0.790 $\pm$ 0.032 & \mbox{\colorbox{gray!35}{0.916 $\pm$ 0.025}} \\[-0.28ex]
		   &  & ECE & 0.048 $\pm$ 0.025 & 0.062 $\pm$ 0.035 & 0.072 $\pm$ 0.037 & 0.056 $\pm$ 0.036 & 0.039 $\pm$ 0.022 & 0.054 $\pm$ 0.029 & 0.050 $\pm$ 0.028 & 0.036 $\pm$ 0.012 & \mbox{\colorbox{gray!35}{0.032 $\pm$ 0.011}} \\[-0.28ex]
		  \cmidrule(lr){2-12}
		   & \multirow{2}{*}{$\mathcal{C}_3$} & Acc. & 0.941 $\pm$ 0.019 & 0.945 $\pm$ 0.018 & 0.906 $\pm$ 0.051 & 0.896 $\pm$ 0.056 & 0.930 $\pm$ 0.022 & 0.946 $\pm$ 0.017 & 0.935 $\pm$ 0.022 & 0.945 $\pm$ 0.011 & \mbox{\colorbox{gray!35}{0.986 $\pm$ 0.013}} \\[-0.28ex]
		   &  & ECE & 0.028 $\pm$ 0.010 & 0.029 $\pm$ 0.010 & 0.028 $\pm$ 0.020 & 0.036 $\pm$ 0.033 & 0.033 $\pm$ 0.012 & 0.027 $\pm$ 0.010 & 0.030 $\pm$ 0.012 & 0.021 $\pm$ 0.006 & \mbox{\colorbox{gray!35}{0.008 $\pm$ 0.007}} \\[-0.28ex]
		  \cmidrule(lr){2-12}
		   & \multirow{2}{*}{\textbf{Average}} & Acc. & 0.808 $\pm$ 0.021 & 0.817 $\pm$ 0.020 & 0.805 $\pm$ 0.019 & 0.798 $\pm$ 0.024 & 0.819 $\pm$ 0.024 & 0.814 $\pm$ 0.021 & 0.808 $\pm$ 0.020 & 0.872 $\pm$ 0.012 & \mbox{\colorbox{gray!35}{0.931 $\pm$ 0.038}} \\[-0.28ex]
		   &  & ECE & 0.058 $\pm$ 0.011 & 0.058 $\pm$ 0.012 & 0.059 $\pm$ 0.027 & 0.062 $\pm$ 0.014 & 0.103 $\pm$ 0.015 & 0.057 $\pm$ 0.011 & 0.056 $\pm$ 0.011 & 0.071 $\pm$ 0.009 & \mbox{\colorbox{gray!35}{0.034 $\pm$ 0.024}} \\[-0.28ex]
			 \cmidrule[0.8pt](r){1-12}
		  \multirow{8}{*}{Teeth Crack} & \multirow{2}{*}{$\mathcal{C}_1$} & Acc. & 0.735 $\pm$ 0.069 & 0.752 $\pm$ 0.070 & 0.800 $\pm$ 0.093 & 0.799 $\pm$ 0.072 & 0.791 $\pm$ 0.077 & 0.743 $\pm$ 0.070 & 0.746 $\pm$ 0.070 & 0.882 $\pm$ 0.040 & \mbox{\colorbox{gray!35}{0.891 $\pm$ 0.082}} \\[-0.28ex]
		   &  & ECE & 0.145 $\pm$ 0.019 & 0.143 $\pm$ 0.021 & 0.191 $\pm$ 0.024 & 0.185 $\pm$ 0.032 & 0.266 $\pm$ 0.055 & 0.145 $\pm$ 0.019 & 0.147 $\pm$ 0.021 & 0.207 $\pm$ 0.020 & \mbox{\colorbox{gray!35}{0.075 $\pm$ 0.070}} \\[-0.28ex]
		  \cmidrule(lr){2-12}
		   & \multirow{2}{*}{$\mathcal{C}_2$} & Acc. & 0.746 $\pm$ 0.016 & 0.740 $\pm$ 0.028 & 0.669 $\pm$ 0.050 & 0.711 $\pm$ 0.025 & 0.761 $\pm$ 0.016 & 0.745 $\pm$ 0.021 & 0.743 $\pm$ 0.020 & \mbox{\colorbox{gray!35}{0.810 $\pm$ 0.020}} & 0.804 $\pm$ 0.040 \\[-0.28ex]
		   &  & ECE & 0.056 $\pm$ 0.020 & \mbox{\colorbox{gray!35}{0.039 $\pm$ 0.016}} & 0.130 $\pm$ 0.019 & 0.075 $\pm$ 0.031 & 0.094 $\pm$ 0.027 & 0.047 $\pm$ 0.017 & 0.054 $\pm$ 0.018 & 0.121 $\pm$ 0.017 & 0.103 $\pm$ 0.028 \\[-0.28ex]
		  \cmidrule(lr){2-12}
		   & \multirow{2}{*}{$\mathcal{C}_3$} & Acc. & \mbox{\colorbox{gray!35}{1.000 $\pm$ 0.000}} & \mbox{\colorbox{gray!35}{1.000 $\pm$ 0.000}} & 0.999 $\pm$ 0.001 & \mbox{\colorbox{gray!35}{1.000 $\pm$ 0.000}} & \mbox{\colorbox{gray!35}{1.000 $\pm$ 0.000}} & \mbox{\colorbox{gray!35}{1.000 $\pm$ 0.000}} & \mbox{\colorbox{gray!35}{1.000 $\pm$ 0.000}} & \mbox{\colorbox{gray!35}{1.000 $\pm$ 0.000}} & 1.000 $\pm$ 0.001 \\[-0.28ex]
		   &  & ECE & 0.002 $\pm$ 0.001 & 0.002 $\pm$ 0.001 & 0.014 $\pm$ 0.008 & 0.005 $\pm$ 0.003 & 0.002 $\pm$ 0.001 & 0.002 $\pm$ 0.001 & 0.002 $\pm$ 0.001 & \mbox{\colorbox{gray!35}{0.001 $\pm$ 0.000}} & 0.002 $\pm$ 0.002 \\[-0.28ex]
		  \cmidrule(lr){2-12}
		   & \multirow{2}{*}{\textbf{Average}} & Acc. & 0.827 $\pm$ 0.025 & 0.831 $\pm$ 0.027 & 0.823 $\pm$ 0.034 & 0.837 $\pm$ 0.023 & 0.850 $\pm$ 0.025 & 0.829 $\pm$ 0.026 & 0.830 $\pm$ 0.023 & 0.897 $\pm$ 0.012 & \mbox{\colorbox{gray!35}{0.898 $\pm$ 0.038}} \\[-0.28ex]
		   &  & ECE & 0.063 $\pm$ 0.012 & \mbox{\colorbox{gray!35}{0.053 $\pm$ 0.012}} & 0.088 $\pm$ 0.019 & 0.076 $\pm$ 0.019 & 0.115 $\pm$ 0.025 & 0.059 $\pm$ 0.012 & 0.061 $\pm$ 0.012 & 0.107 $\pm$ 0.008 & 0.058 $\pm$ 0.025 \\[-0.28ex]
		  \cmidrule[0.8pt](r){1-12}
		  \multicolumn{3}{c}{\emph{Total Average}} & 0.815 $\pm$ 0.025 & 0.818 $\pm$ 0.026 & 0.823 $\pm$ 0.030 & 0.820 $\pm$ 0.023 & 0.832 $\pm$ 0.026 & 0.817 $\pm$ 0.025 & 0.816 $\pm$ 0.024 & 0.881 $\pm$ 0.013 & \mbox{\colorbox{gray!35}{0.918 $\pm$ 0.037}} \\
		  \cmidrule[0.8pt](r){1-12}
		  \multicolumn{3}{c}{\emph{Total Rank}} & 9 & 6 & 4 & 5 & 3 & 7 & 8 & 2 & \textbf{1} \\
		  \specialrule{0.1em}{2pt}{1pt}
		  \specialrule{0.1em}{1pt}{1pt}
	  \end{tabular}%
	  }
	  \begin{tablenotes}
		\footnotesize
		\item[]Notes: The best results are highlighted with \raisebox{0.35ex}{\mbox{\colorbox{gray!35}{\rule{0pt}{0.2em}\rule{0.2em}{0pt}}}} gray shading.
	  \end{tablenotes}
	  \end{threeparttable}
	\end{center}
\end{table*}

\begin{figure*}[htbp]
	\centering
	\subfloat[Gear wear, Cumulative]{\includegraphics[width=0.33\textwidth]{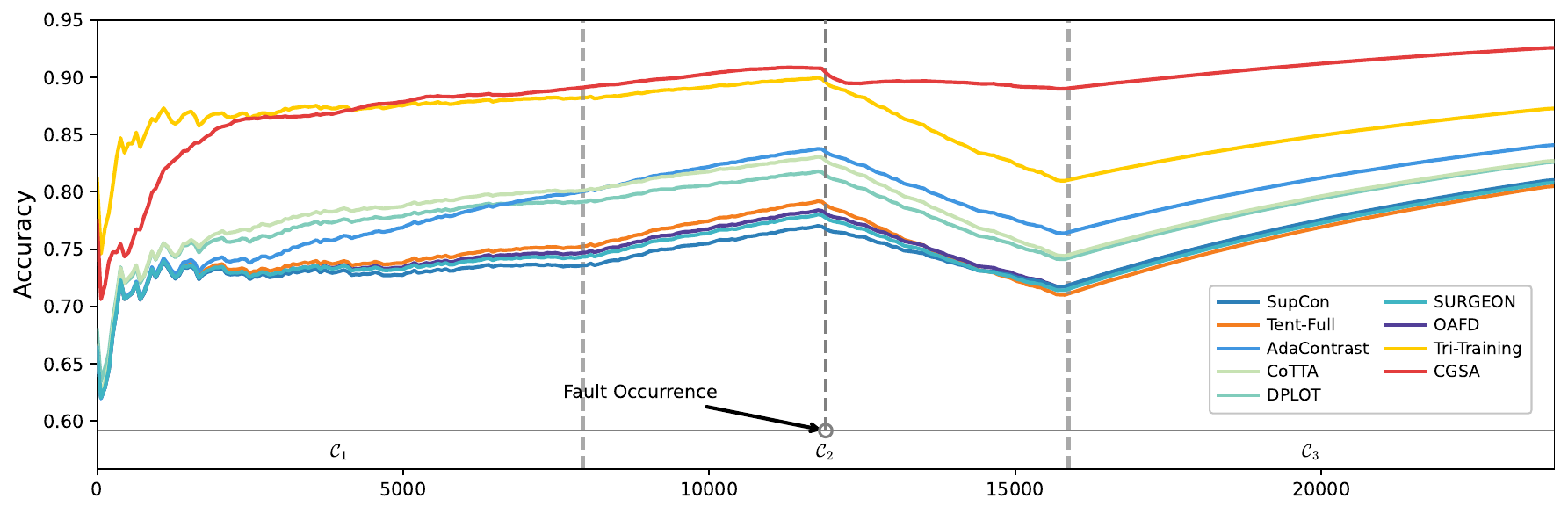}\label{Gear wear, Cumulative}}
	\hfil
	\subfloat[Teeth break, Cumulative]{\includegraphics[width=0.33\textwidth]{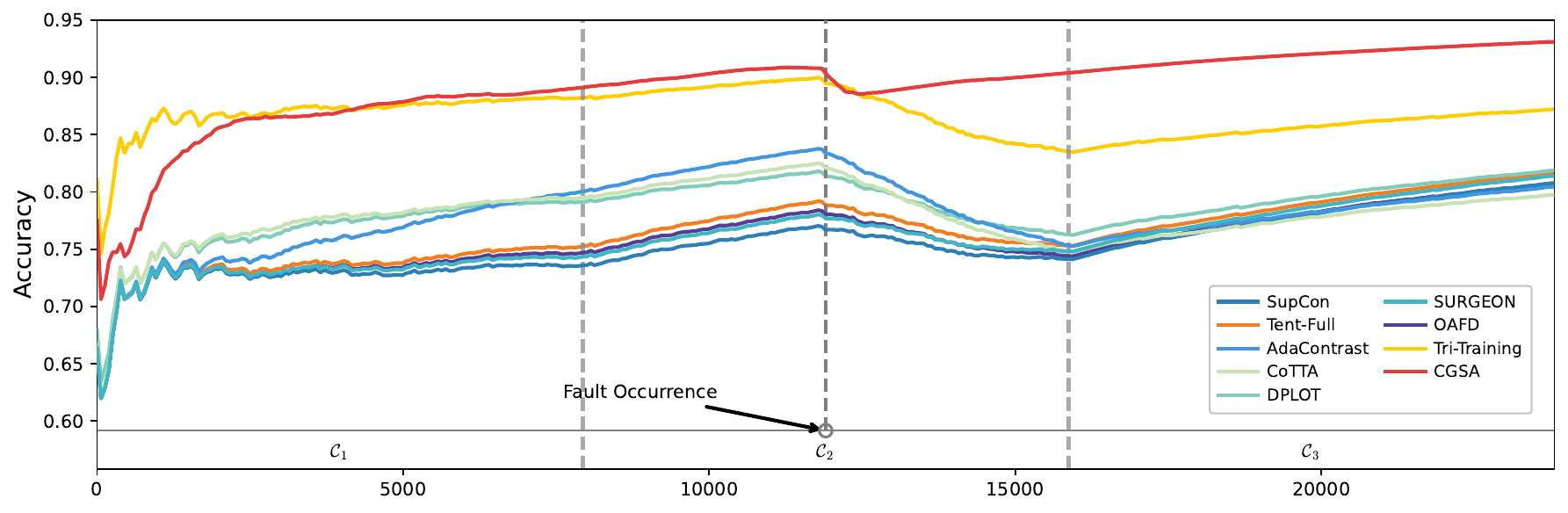}\label{Teeth break, Cumulative}}
	\hfil
	\subfloat[Teeth crack, Cumulative]{\includegraphics[width=0.33\textwidth]{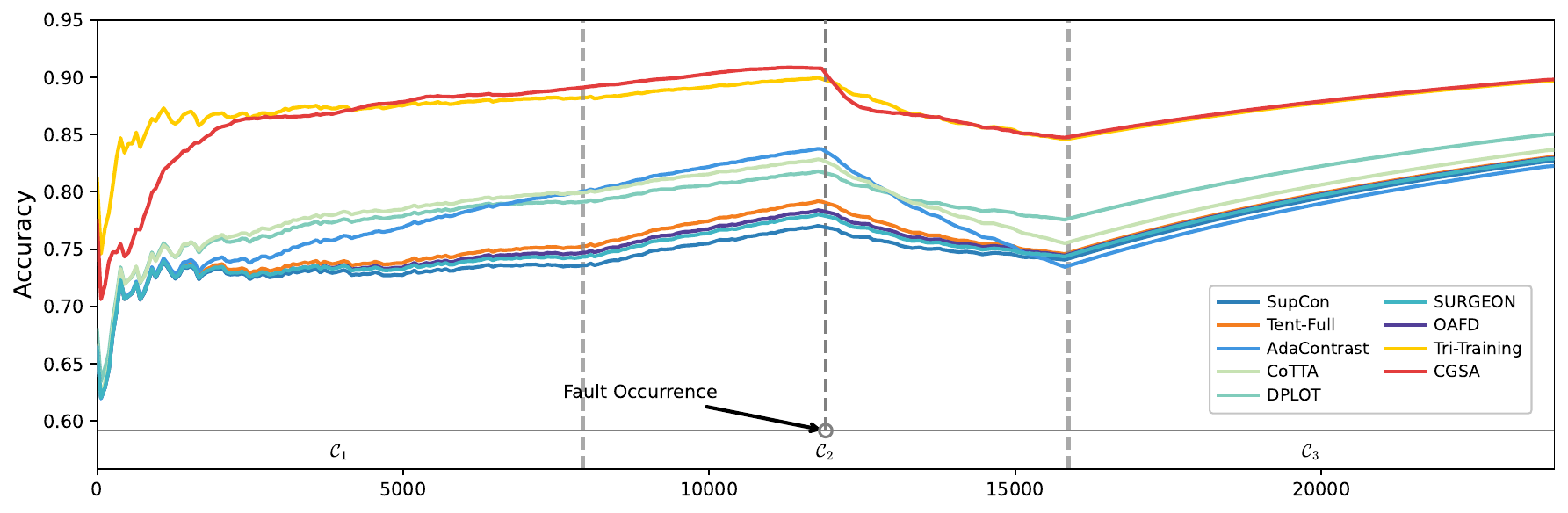}\label{Teeth crack, Cumulative}}\\
	\subfloat[Gear wear, Real-time]{\includegraphics[width=0.33\textwidth]{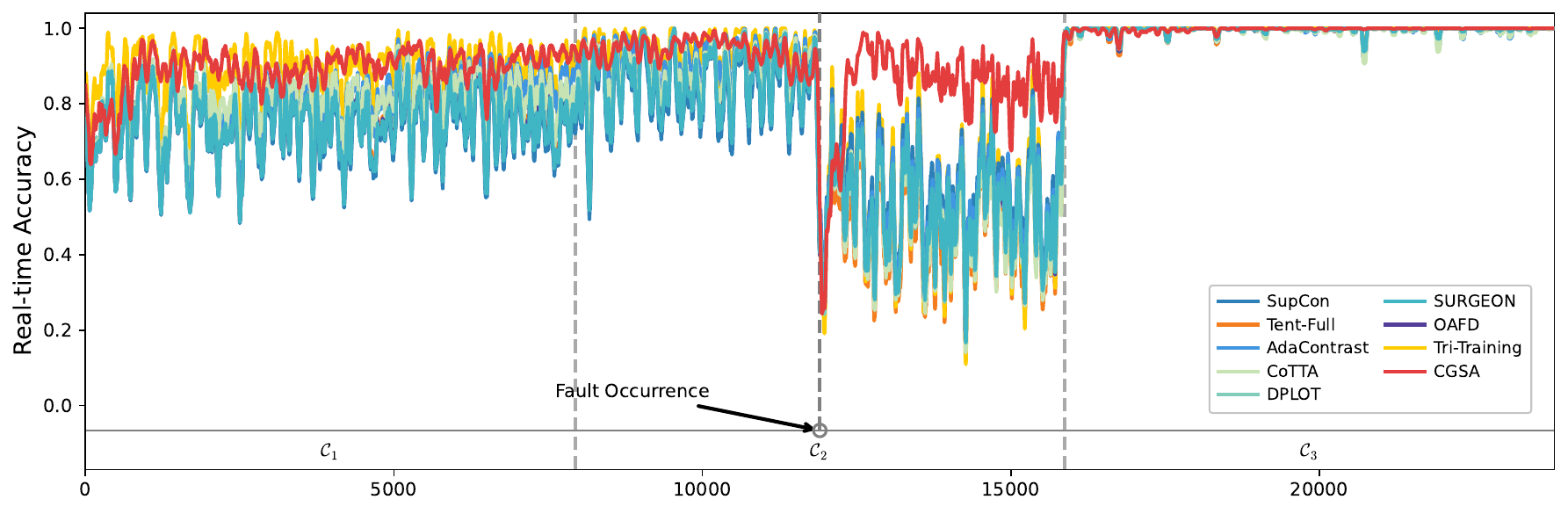}\label{Gear wear, Real-time}}
	\hfil
	\subfloat[Teeth break, Real-time]{\includegraphics[width=0.33\textwidth]{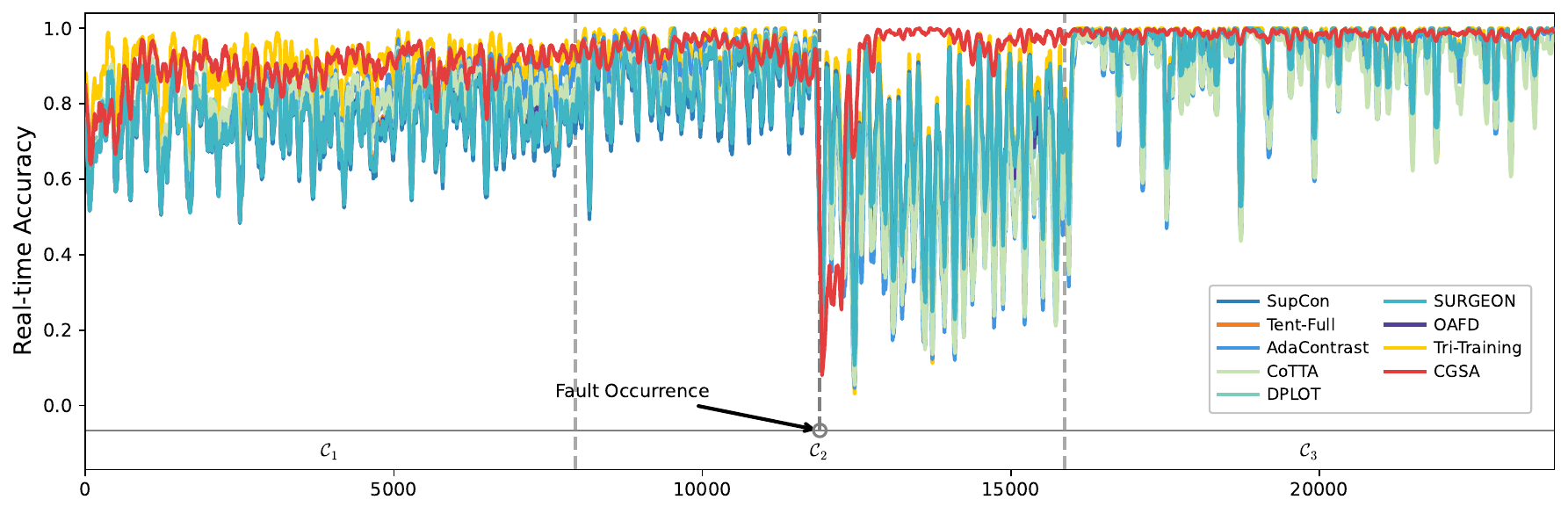}\label{Teeth break, Real-time}}
	\hfil
	\subfloat[Teeth crack, Real-time]{\includegraphics[width=0.33\textwidth]{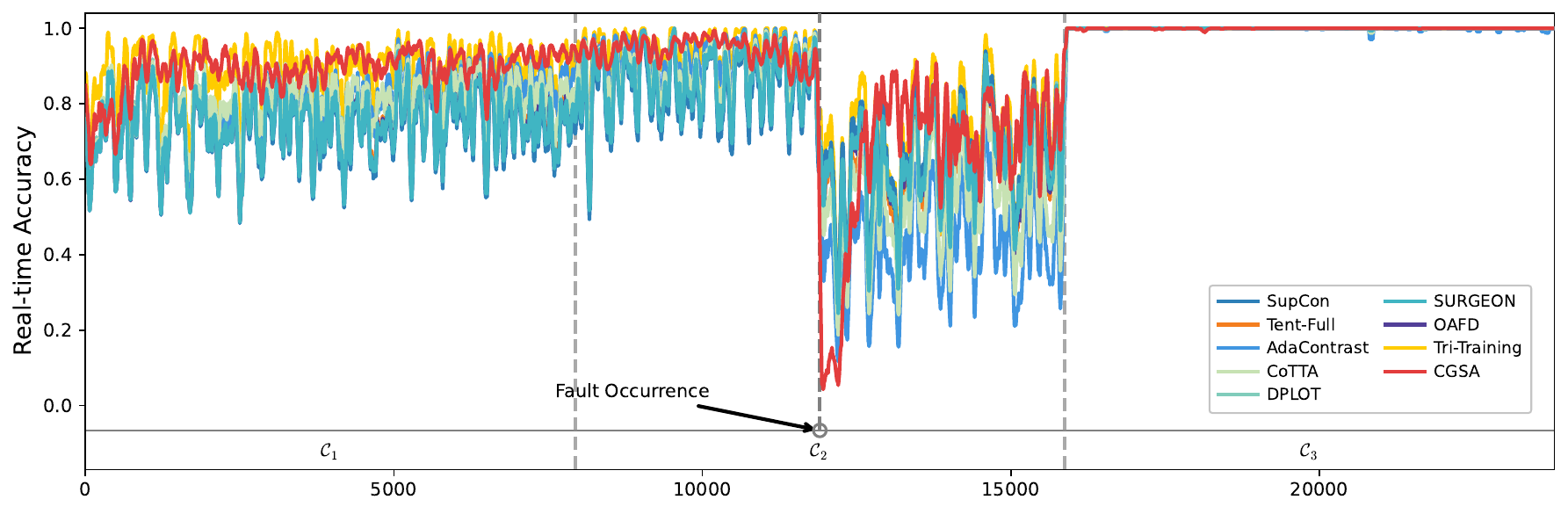}\label{Teeth crack, Real-time}}
	\caption{Diagnostic performance of different methods on the gearbox dataset.}
\end{figure*}

Moreover, all methods use a three-layer MLP 
as the feature extractor and a fully connected layer as the fault classifier.
To enable adversarial learning, a domain discriminator is introduced. 
In DANN, the condition discriminator is implemented as a four-layer MLP. 
In contrast, the proposed regression discriminator adopts a dual-branch architecture, 
where each branch is also implemented as a four-layer MLP to estimate the mean and variance, respectively.
In the offline stage, methods are trained for 200 epochs to ensure convergence.
For the proposed method, the progressive increasing rate $\gamma$ is set to 10.
In the online stage, all methods are evaluated with an online batch size of $B_{\text{on}}=256$. 
For the proposed method, the Adam optimizer is adopted with a learning rate of $5 \times 10^{-4}$. 
The queue length \(W\) is set to 50. 
The residual condition consistency $\tau_{\mu}$ is set to 0.05, 
and the pseudo-label confidence threshold $\epsilon_p$ is set to 0.9.

The proposed method is compared with several representative approaches from different perspectives, 
including supervised learning, online test-time adaptation, and online semi-supervised learning.
Specifically, SupCon \cite{SupCon} is adopted as a representative supervised learning method, 
Tent-Full \cite{Tent}, AdaContrast \cite{AdaContrast}, CoTTA \cite{CoTTA}, DPLOT \cite{DPLOT}, SURGEON \cite{SURGEON}, and OAFD \cite{OTTA-instance-2} are selected as advanced online test-time adaptation methods;
and Tri-Training \cite{Tri-training} is selected as an online semi-supervised learning method.
\color{black}
Here, Tent-Full denotes a variant of Tent that updates all parameters rather than only the BN layers.
For a fair comparison, all online methods are initialized from the same source pre-trained model obtained by offline SupCon training, 
and the hyperparameters of all compared methods are tuned under the same validation protocol for each dataset to ensure optimal performance.
The proposed \MethodName, adopting a complete offline and online framework, is compared with all the methods mentioned above under the same experimental settings.
\color{black}

\begin{table*}[!t]
	\begin{center}
	  \centering
	  \caption{The comparison results on motor Dataset regarding average Acc.($\uparrow$) and ECE($\downarrow$).}\label{table_comparison_motor}
	  \begin{threeparttable}
	  \footnotesize
	  \setlength{\tabcolsep}{1.0mm}
	  \resizebox{\textwidth}{!}{%
	  \begin{tabular}{c|c|c|ccccccccc}
		  \specialrule{0.1em}{1pt}{1pt}
		  \specialrule{0.1em}{1pt}{2pt}
		  \cmidrulewidth = 0.5em
		  \multirow{2.5}{*}{Fault Type} & \multirow{2.5}{*}{Stage} & \multirow{2.5}{*}{Metric} & \multicolumn{9}{c}{Methods} \\
		  \cmidrule(lr){4-12}
		  &  &  & \multicolumn{1}{c}{\textit{SupCon}} & \multicolumn{1}{c}{\textit{Tent-Full}} & \multicolumn{1}{c}{\textit{AdaContrast}} & \multicolumn{1}{c}{\textit{CoTTA}} & \multicolumn{1}{c}{\textit{DPLOT}} & \multicolumn{1}{c}{\textit{SURGEON}} & \multicolumn{1}{c}{\textit{OAFD}} & \multicolumn{1}{c}{\textit{Tri-Training}} & \multicolumn{1}{c}{\textit{CGSA}\textsuperscript{*}} \\
		\specialrule{0.10em}{2pt}{2pt}
		  \renewcommand{\arraystretch}{0.9}
		  \multirow{8}{*}{Voltage Unbalance} & \multirow{2}{*}{$\mathcal{C}_1$} & Acc. & 0.760 $\pm$ 0.059 & 0.846 $\pm$ 0.080 & 0.760 $\pm$ 0.061 & 0.817 $\pm$ 0.053 & 0.850 $\pm$ 0.087 & 0.850 $\pm$ 0.079 & 0.788 $\pm$ 0.058 & 0.854 $\pm$ 0.039 & \mbox{\colorbox{gray!35}{0.881 $\pm$ 0.092}} \\[-0.28ex]
		   &  & ECE & 0.147 $\pm$ 0.055 & 0.100 $\pm$ 0.073 & 0.147 $\pm$ 0.056 & 0.089 $\pm$ 0.048 & 0.087 $\pm$ 0.074 & 0.097 $\pm$ 0.073 & 0.121 $\pm$ 0.052 & \mbox{\colorbox{gray!35}{0.062 $\pm$ 0.032}} & 0.076 $\pm$ 0.075 \\[-0.28ex]
		  \cmidrule(lr){2-12}
		   & \multirow{2}{*}{$\mathcal{C}_2$} & Acc. & 0.371 $\pm$ 0.045 & 0.449 $\pm$ 0.069 & 0.372 $\pm$ 0.047 & 0.475 $\pm$ 0.026 & 0.457 $\pm$ 0.063 & 0.454 $\pm$ 0.066 & 0.419 $\pm$ 0.035 & 0.433 $\pm$ 0.022 & \mbox{\colorbox{gray!35}{0.718 $\pm$ 0.055}} \\[-0.28ex]
		   &  & ECE & 0.545 $\pm$ 0.047 & 0.535 $\pm$ 0.061 & 0.546 $\pm$ 0.050 & 0.472 $\pm$ 0.024 & 0.531 $\pm$ 0.057 & 0.531 $\pm$ 0.059 & 0.498 $\pm$ 0.036 & 0.495 $\pm$ 0.018 & \mbox{\colorbox{gray!35}{0.228 $\pm$ 0.058}} \\[-0.28ex]
		  \cmidrule(lr){2-12}
		   & \multirow{2}{*}{$\mathcal{C}_3$} & Acc. & \mbox{\colorbox{gray!35}{1.000 $\pm$ 0.000}} & \mbox{\colorbox{gray!35}{1.000 $\pm$ 0.000}} & \mbox{\colorbox{gray!35}{1.000 $\pm$ 0.000}} & \mbox{\colorbox{gray!35}{1.000 $\pm$ 0.000}} & \mbox{\colorbox{gray!35}{1.000 $\pm$ 0.000}} & \mbox{\colorbox{gray!35}{1.000 $\pm$ 0.000}} & \mbox{\colorbox{gray!35}{1.000 $\pm$ 0.000}} & \mbox{\colorbox{gray!35}{1.000 $\pm$ 0.000}} & 0.997 $\pm$ 0.003 \\[-0.28ex]
		   &  & ECE & \mbox{\colorbox{gray!35}{0.000 $\pm$ 0.000}} & \mbox{\colorbox{gray!35}{0.000 $\pm$ 0.000}} & \mbox{\colorbox{gray!35}{0.000 $\pm$ 0.000}} & 0.000 $\pm$ 0.001 & \mbox{\colorbox{gray!35}{0.000 $\pm$ 0.000}} & \mbox{\colorbox{gray!35}{0.000 $\pm$ 0.000}} & 0.000 $\pm$ 0.001 & \mbox{\colorbox{gray!35}{0.000 $\pm$ 0.000}} & 0.003 $\pm$ 0.003 \\[-0.28ex]
		  \cmidrule(lr){2-12}
		   & \multirow{2}{*}{\textbf{Average}} & Acc. & 0.710 $\pm$ 0.032 & 0.765 $\pm$ 0.049 & 0.711 $\pm$ 0.034 & 0.764 $\pm$ 0.024 & 0.769 $\pm$ 0.049 & 0.768 $\pm$ 0.048 & 0.736 $\pm$ 0.029 & 0.762 $\pm$ 0.020 & \mbox{\colorbox{gray!35}{0.865 $\pm$ 0.044}} \\[-0.28ex]
		   &  & ECE & 0.231 $\pm$ 0.032 & 0.211 $\pm$ 0.044 & 0.231 $\pm$ 0.034 & 0.186 $\pm$ 0.023 & 0.205 $\pm$ 0.044 & 0.209 $\pm$ 0.043 & 0.206 $\pm$ 0.027 & 0.185 $\pm$ 0.017 & \mbox{\colorbox{gray!35}{0.101 $\pm$ 0.041}} \\[-0.28ex]
			 \cmidrule[0.8pt](r){1-12}
		  \multirow{8}{*}{Static Eccentricity} & \multirow{2}{*}{$\mathcal{C}_1$} & Acc. & 0.760 $\pm$ 0.059 & 0.846 $\pm$ 0.080 & 0.760 $\pm$ 0.061 & 0.818 $\pm$ 0.054 & 0.850 $\pm$ 0.087 & 0.850 $\pm$ 0.079 & 0.788 $\pm$ 0.058 & 0.854 $\pm$ 0.039 & \mbox{\colorbox{gray!35}{0.881 $\pm$ 0.092}} \\[-0.28ex]
		   &  & ECE & 0.147 $\pm$ 0.055 & 0.100 $\pm$ 0.073 & 0.147 $\pm$ 0.056 & 0.089 $\pm$ 0.049 & 0.087 $\pm$ 0.074 & 0.097 $\pm$ 0.073 & 0.121 $\pm$ 0.052 & \mbox{\colorbox{gray!35}{0.062 $\pm$ 0.032}} & 0.076 $\pm$ 0.075 \\[-0.28ex]
		  \cmidrule(lr){2-12}
		   & \multirow{2}{*}{$\mathcal{C}_2$} & Acc. & 0.843 $\pm$ 0.045 & 0.938 $\pm$ 0.072 & 0.843 $\pm$ 0.048 & \mbox{\colorbox{gray!35}{0.952 $\pm$ 0.030}} & 0.947 $\pm$ 0.068 & 0.944 $\pm$ 0.069 & 0.894 $\pm$ 0.033 & 0.915 $\pm$ 0.024 & 0.926 $\pm$ 0.057 \\[-0.28ex]
		   &  & ECE & 0.085 $\pm$ 0.049 & 0.052 $\pm$ 0.065 & 0.086 $\pm$ 0.052 & \mbox{\colorbox{gray!35}{0.023 $\pm$ 0.018}} & 0.046 $\pm$ 0.063 & 0.047 $\pm$ 0.062 & 0.039 $\pm$ 0.030 & 0.024 $\pm$ 0.019 & 0.044 $\pm$ 0.047 \\[-0.28ex]
		  \cmidrule(lr){2-12}
		   & \multirow{2}{*}{$\mathcal{C}_3$} & Acc. & \mbox{\colorbox{gray!35}{1.000 $\pm$ 0.000}} & \mbox{\colorbox{gray!35}{1.000 $\pm$ 0.000}} & \mbox{\colorbox{gray!35}{1.000 $\pm$ 0.000}} & \mbox{\colorbox{gray!35}{1.000 $\pm$ 0.000}} & \mbox{\colorbox{gray!35}{1.000 $\pm$ 0.000}} & \mbox{\colorbox{gray!35}{1.000 $\pm$ 0.000}} & \mbox{\colorbox{gray!35}{1.000 $\pm$ 0.000}} & \mbox{\colorbox{gray!35}{1.000 $\pm$ 0.000}} & \mbox{\colorbox{gray!35}{1.000 $\pm$ 0.000}} \\[-0.28ex]
		   &  & ECE & \mbox{\colorbox{gray!35}{0.000 $\pm$ 0.000}} & \mbox{\colorbox{gray!35}{0.000 $\pm$ 0.000}} & \mbox{\colorbox{gray!35}{0.000 $\pm$ 0.000}} & \mbox{\colorbox{gray!35}{0.000 $\pm$ 0.000}} & \mbox{\colorbox{gray!35}{0.000 $\pm$ 0.000}} & \mbox{\colorbox{gray!35}{0.000 $\pm$ 0.000}} & \mbox{\colorbox{gray!35}{0.000 $\pm$ 0.000}} & \mbox{\colorbox{gray!35}{0.000 $\pm$ 0.000}} & \mbox{\colorbox{gray!35}{0.000 $\pm$ 0.000}} \\[-0.28ex]
		  \cmidrule(lr){2-12}
		   & \multirow{2}{*}{\textbf{Average}} & Acc. & 0.867 $\pm$ 0.033 & 0.928 $\pm$ 0.050 & 0.868 $\pm$ 0.034 & 0.923 $\pm$ 0.026 & 0.932 $\pm$ 0.051 & 0.932 $\pm$ 0.049 & 0.894 $\pm$ 0.029 & 0.923 $\pm$ 0.021 & \mbox{\colorbox{gray!35}{0.936 $\pm$ 0.032}} \\[-0.28ex]
		   &  & ECE & 0.077 $\pm$ 0.032 & 0.050 $\pm$ 0.046 & 0.077 $\pm$ 0.034 & 0.032 $\pm$ 0.024 & 0.044 $\pm$ 0.045 & 0.047 $\pm$ 0.045 & 0.052 $\pm$ 0.027 & \mbox{\colorbox{gray!35}{0.028 $\pm$ 0.018}} & 0.040 $\pm$ 0.027 \\[-0.28ex]
			 \cmidrule[0.8pt](r){1-12}
		  \multirow{8}{*}{Bearing Ball Fault} & \multirow{2}{*}{$\mathcal{C}_1$} & Acc. & 0.760 $\pm$ 0.059 & 0.846 $\pm$ 0.080 & 0.760 $\pm$ 0.061 & 0.819 $\pm$ 0.054 & 0.850 $\pm$ 0.087 & 0.850 $\pm$ 0.079 & 0.788 $\pm$ 0.058 & 0.854 $\pm$ 0.039 & \mbox{\colorbox{gray!35}{0.881 $\pm$ 0.092}} \\[-0.28ex]
		   &  & ECE & 0.147 $\pm$ 0.055 & 0.100 $\pm$ 0.073 & 0.147 $\pm$ 0.056 & 0.089 $\pm$ 0.049 & 0.087 $\pm$ 0.074 & 0.097 $\pm$ 0.073 & 0.121 $\pm$ 0.052 & \mbox{\colorbox{gray!35}{0.062 $\pm$ 0.032}} & 0.076 $\pm$ 0.075 \\[-0.28ex]
		  \cmidrule(lr){2-12}
		   & \multirow{2}{*}{$\mathcal{C}_2$} & Acc. & 0.793 $\pm$ 0.056 & 0.531 $\pm$ 0.040 & 0.791 $\pm$ 0.061 & 0.578 $\pm$ 0.029 & 0.528 $\pm$ 0.033 & 0.525 $\pm$ 0.033 & 0.769 $\pm$ 0.050 & 0.842 $\pm$ 0.031 & \mbox{\colorbox{gray!35}{0.877 $\pm$ 0.048}} \\[-0.28ex]
		   &  & ECE & 0.088 $\pm$ 0.078 & 0.448 $\pm$ 0.052 & 0.093 $\pm$ 0.084 & 0.303 $\pm$ 0.048 & 0.458 $\pm$ 0.045 & 0.456 $\pm$ 0.046 & 0.091 $\pm$ 0.068 & \mbox{\colorbox{gray!35}{0.039 $\pm$ 0.023}} & 0.064 $\pm$ 0.045 \\[-0.28ex]
		  \cmidrule(lr){2-12}
		   & \multirow{2}{*}{$\mathcal{C}_3$} & Acc. & \mbox{\colorbox{gray!35}{1.000 $\pm$ 0.000}} & 0.999 $\pm$ 0.004 & \mbox{\colorbox{gray!35}{1.000 $\pm$ 0.000}} & 0.998 $\pm$ 0.007 & 0.998 $\pm$ 0.005 & 0.999 $\pm$ 0.004 & \mbox{\colorbox{gray!35}{1.000 $\pm$ 0.000}} & \mbox{\colorbox{gray!35}{1.000 $\pm$ 0.000}} & \mbox{\colorbox{gray!35}{1.000 $\pm$ 0.000}} \\[-0.28ex]
		   &  & ECE & \mbox{\colorbox{gray!35}{0.000 $\pm$ 0.000}} & 0.002 $\pm$ 0.002 & \mbox{\colorbox{gray!35}{0.000 $\pm$ 0.000}} & 0.004 $\pm$ 0.005 & 0.002 $\pm$ 0.003 & 0.002 $\pm$ 0.002 & 0.001 $\pm$ 0.001 & \mbox{\colorbox{gray!35}{0.000 $\pm$ 0.000}} & 0.001 $\pm$ 0.001 \\[-0.28ex]
		  \cmidrule(lr){2-12}
		   & \multirow{2}{*}{\textbf{Average}} & Acc. & 0.851 $\pm$ 0.034 & 0.792 $\pm$ 0.017 & 0.850 $\pm$ 0.036 & 0.798 $\pm$ 0.020 & 0.792 $\pm$ 0.023 & 0.791 $\pm$ 0.018 & 0.852 $\pm$ 0.032 & 0.899 $\pm$ 0.021 & \mbox{\colorbox{gray!35}{0.919 $\pm$ 0.033}} \\[-0.28ex]
		   &  & ECE & 0.078 $\pm$ 0.041 & 0.182 $\pm$ 0.015 & 0.079 $\pm$ 0.044 & 0.130 $\pm$ 0.023 & 0.181 $\pm$ 0.019 & 0.184 $\pm$ 0.015 & 0.066 $\pm$ 0.039 & \mbox{\colorbox{gray!35}{0.030 $\pm$ 0.021}} & 0.046 $\pm$ 0.029 \\[-0.28ex]
		  \cmidrule[0.8pt](r){1-12}
		  \multicolumn{3}{c}{\emph{Total Average}} & 0.810 $\pm$ 0.033 & 0.828 $\pm$ 0.039 & 0.810 $\pm$ 0.035 & 0.829 $\pm$ 0.023 & 0.831 $\pm$ 0.041 & 0.830 $\pm$ 0.038 & 0.827 $\pm$ 0.030 & 0.861 $\pm$ 0.021 & \mbox{\colorbox{gray!35}{0.907 $\pm$ 0.036}} \\
		  \cmidrule[0.8pt](r){1-12}
		  \multicolumn{3}{c}{\emph{Total Rank}} & 8 & 6 & 9 & 5 & 3 & 4 & 7 & 2 & \textbf{1} \\
		  \specialrule{0.1em}{2pt}{1pt}
		  \specialrule{0.1em}{1pt}{1pt}
	  \end{tabular}%
	  }
	  \begin{tablenotes}
		\footnotesize
		\item[]Notes: The best results are highlighted with \raisebox{0.35ex}{\mbox{\colorbox{gray!35}{\rule{0pt}{0.2em}\rule{0.2em}{0pt}}}} gray shading.
	  \end{tablenotes}
	  \end{threeparttable}
	\end{center}
\end{table*}

\begin{figure*}[htbp]
	\centering
	\subfloat[Voltage Unbalance, Cumulative]{\includegraphics[width=0.33\textwidth]{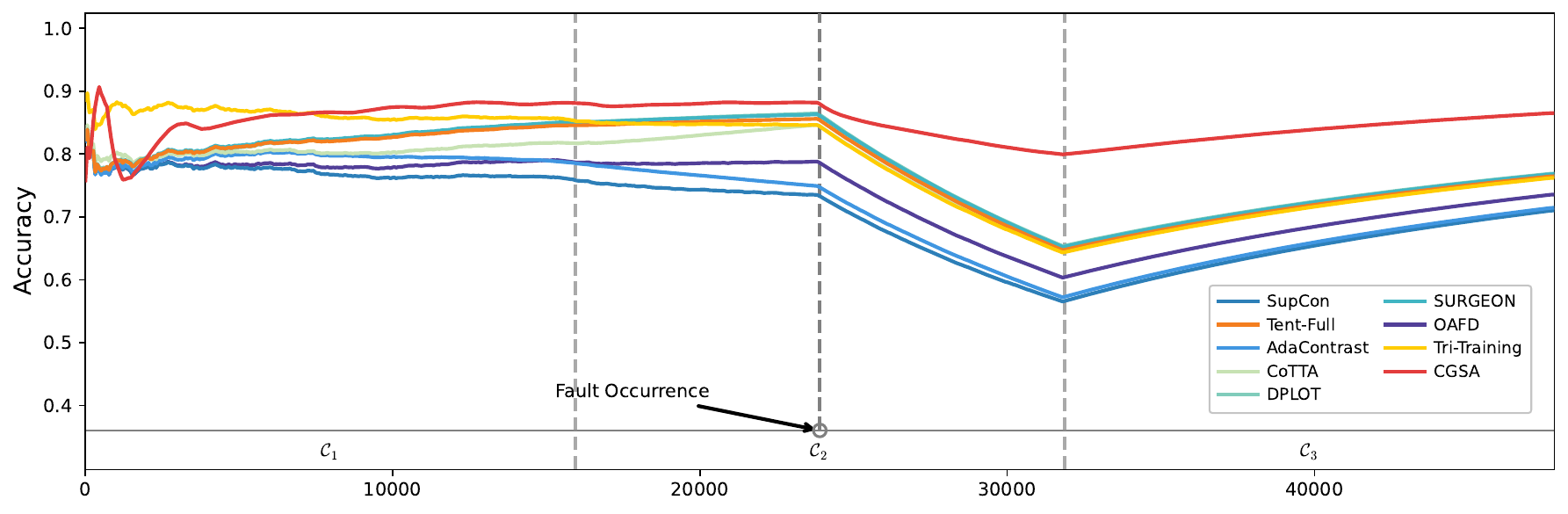}\label{Voltage Unbalance, Cumulative}}
	\hfil
	\subfloat[Static Eccentricity, Cumulative]{\includegraphics[width=0.33\textwidth]{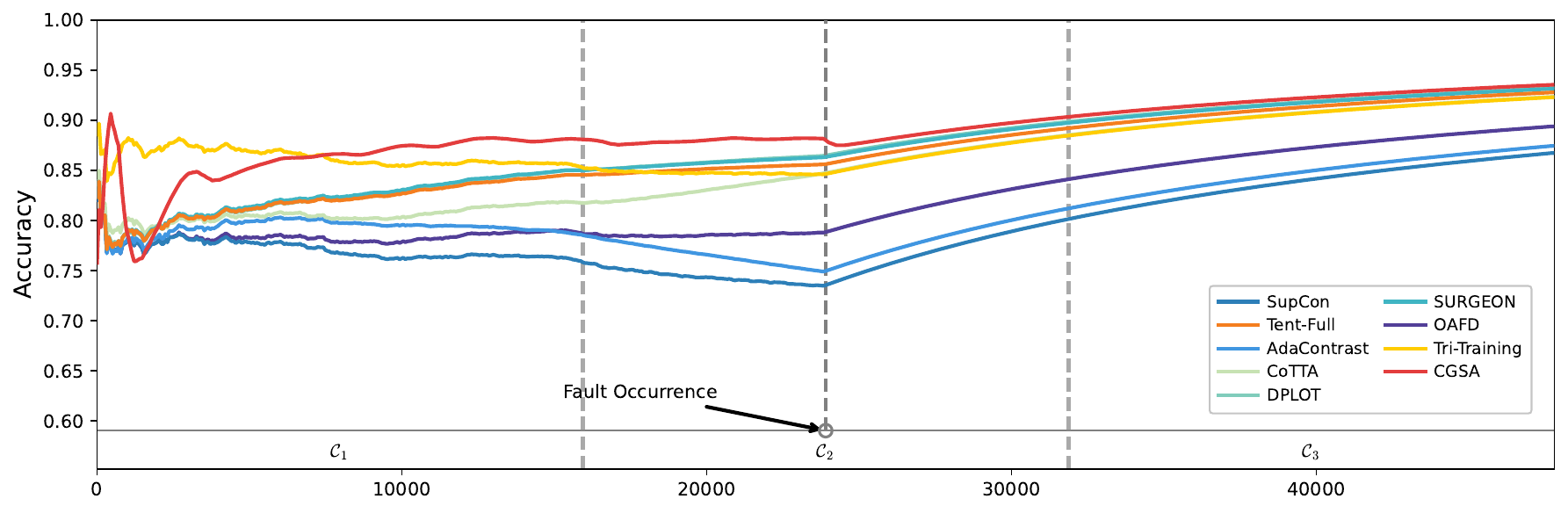}\label{Static Eccentricity, Cumulative}}
	\hfil
	\subfloat[Bearing Ball Fault, Cumulative]{\includegraphics[width=0.33\textwidth]{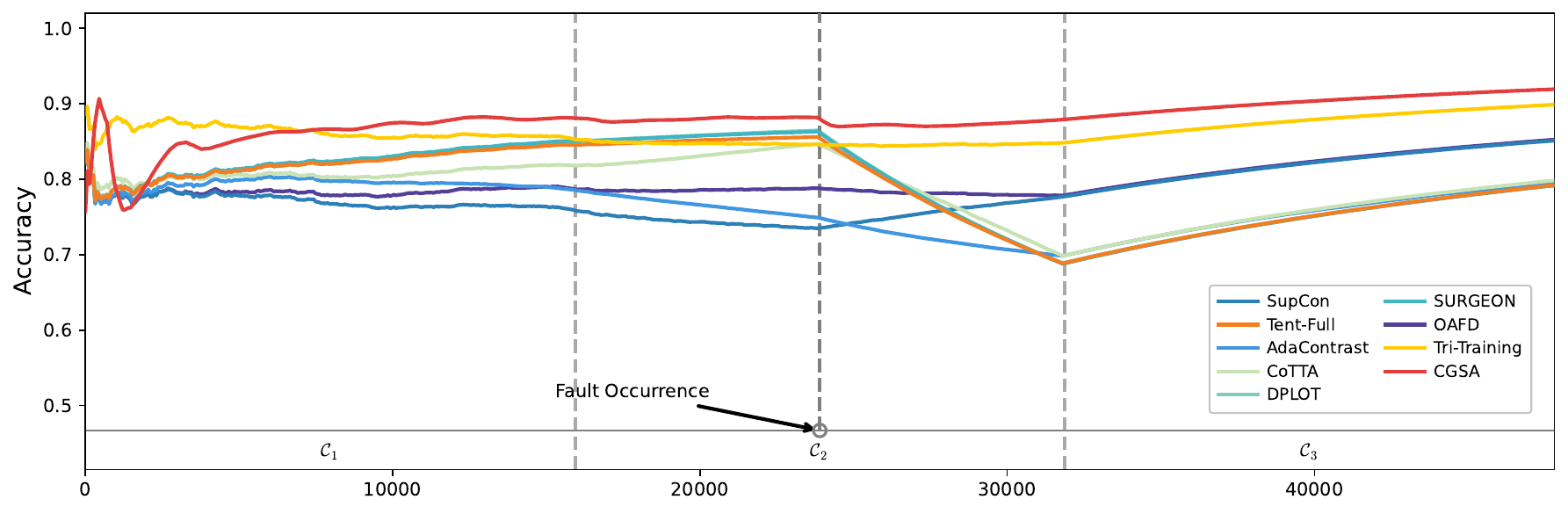}\label{Bearing Ball, Cumulative}} \\
	\subfloat[Voltage Unbalance, Real-time]{\includegraphics[width=0.33\textwidth]{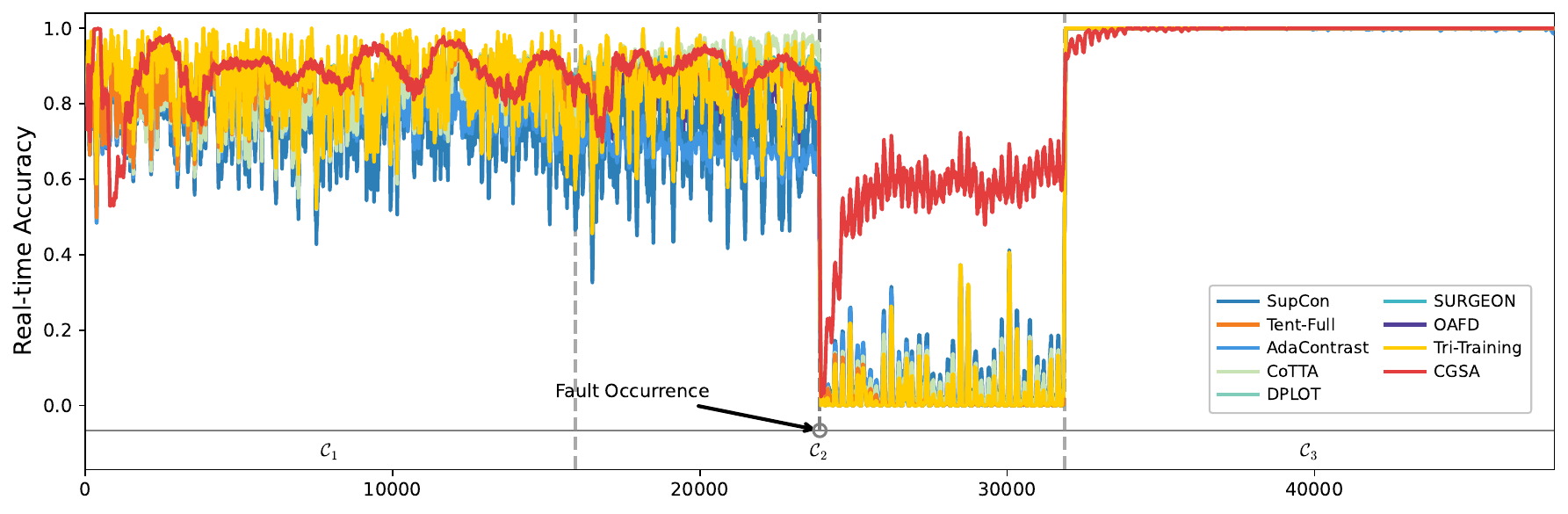}\label{Voltage Unbalance, Real-time}}
	\hfil
	\subfloat[Static Eccentricity, Real-time]{\includegraphics[width=0.33\textwidth]{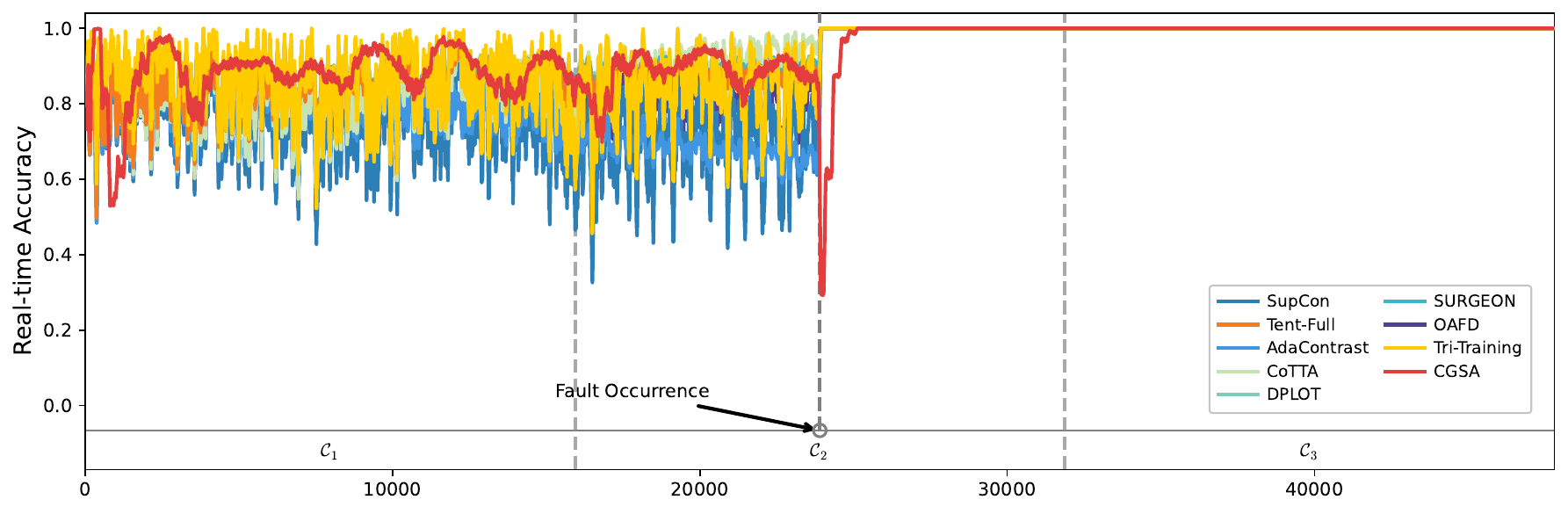}\label{Static Eccentricity, Real-time}}
	\hfil
	\subfloat[Bearing Ball Fault, Real-time]{\includegraphics[width=0.33\textwidth]{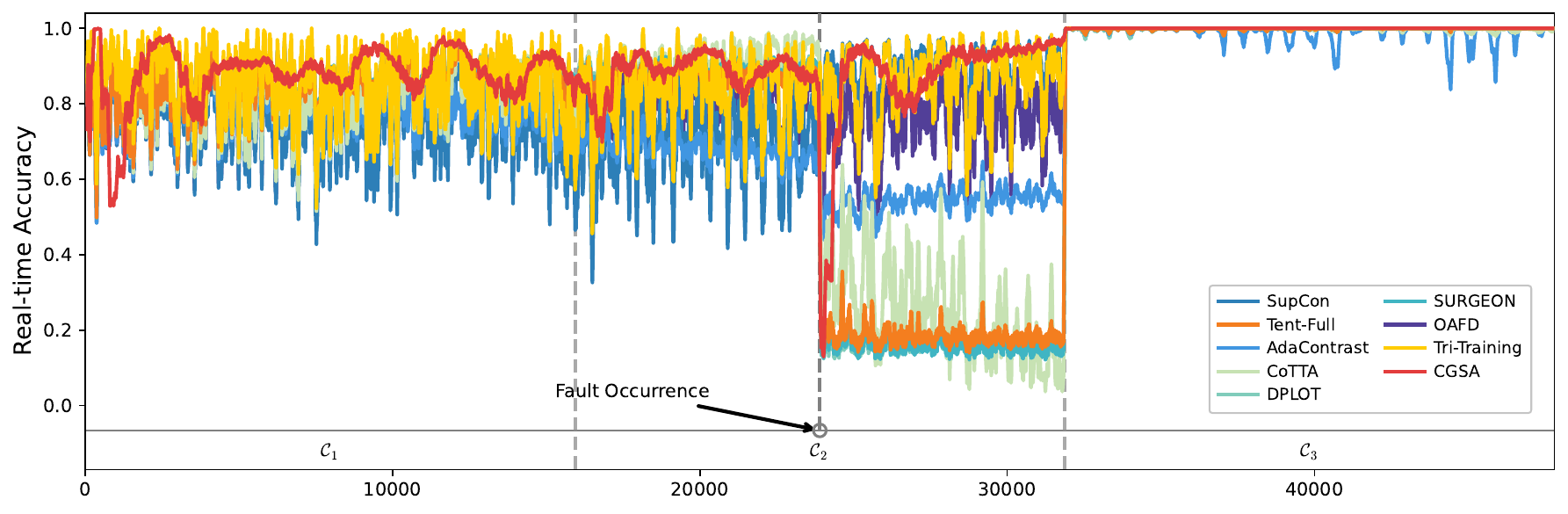}\label{Bearing Ball, Real-time}}
	\caption{Diagnostic performance of different methods on the motor dataset.}
\end{figure*}

\subsection{Case 1: Parallel Gearbox}
\subsubsection{Experimental Settings}
The gearbox test rig (as illustrated in Fig. \ref{fig:mcc5-thu-gearbox}) is composed of a 2.2 kW three-phase asynchronous motor, 
a torque sensor, a two-stage parallel gearbox, a magnetic powder brake, a measurement and a control system. 
Two TES001V three-axis vibration sensors (sensitivity: 100 mV/g)
were used to measure vibrations along the x-, y-, and z-axes of the motor and the gearbox shaft at a 12.8 kHz sampling frequency.

\subsubsection{Comparison Study}
The cumulative accuracy curves and real-time accuracy curves based on the MCC5-THU Gearbox dataset are presented 
in Fig.~\ref{Gear wear, Cumulative}--\ref{Teeth crack, Cumulative}, Fig.~\ref{Gear wear, Real-time}--\ref{Teeth crack, Real-time}, respectively.
In Table~\ref{table_comparison_gearbox}, the Acc. and ECE for each operating condition as well as the overall performance are reported.
The proposed method significantly outperforms other comparative methods.
Especially for teeth break fault, our method achieves a diagnostic accuracy of 0.931, 
surpassing Tri-Training by 5.9 \% and the compared OTTA methods by more than 11 \%,
and maintains stable performance in the in-distribution stage, 
demonstrating its robustness against catastrophic forgetting during online adaptation.

\subsection{Case 2: Three-phase Asynchronous Motor}
\subsubsection{Experimental Settings}
The motor test rig (as illustrated in Fig. \ref{fig:mcc5-thu-motor}) consists of a 2.2 kW three-phase asynchronous motor, 
a torque sensor, a two-stage parallel gearbox and a magnetic powder brake. A tri-axial vibration acceleration 
sensor was mounted on the motor drive end. Vibration, three-phase current (measured using current clamps model Fluke-i30s with a sensitivity 100 mV/A), 
torque, and key-phase signals were acquired at a sampling frequency of 12.8 kHz via 8 synchronous channels.

\subsubsection{Comparison Study}
The cumulative accuracy curves and real-time accuracy curves based on the MCC5-THU Motor dataset are presented 
in Fig.~\ref{Voltage Unbalance, Cumulative}--\ref{Bearing Ball, Cumulative}, Fig.~\ref{Voltage Unbalance, Real-time}--\ref{Bearing Ball, Real-time}, respectively.
In Table~\ref{table_comparison_motor}, the Acc. and ECE for each operating condition as well as the overall performance are reported.
In this scenario, classification becomes more challenging due to the similarity between healthy and faulty samples and the significant distribution discrepancy across different operating conditions.
Nevertheless, the proposed method still achieves the highest overall accuracy of 0.907.
When the voltage unbalance fault and bearing ball fault are introduced, 
all comparison methods suffer from a significant accuracy degradation. 
By contrast, only the proposed \MethodName\ can rapidly adjust to the changed data distribution 
and recover to a high accuracy level.

\subsection{Discussion}
\subsubsection{Performance Comparison}
The comparison results verify the effectiveness of the proposed method from several aspects. 
Compared with the offline method SupCon, the proposed method can further adapt to the evolving online stream to enhance its performance. 
Compared with OTTA methods, which tend to suffer from error accumulation, 
the proposed online mechanism provides a more stable adaptation process and improves diagnostic accuracy.
Tri-Training is additionally included as a source-available reference, since online semi-supervised methods are allowed to access source-domain data during the online stage.
In contrast, the proposed \MethodName\ strictly follows the source-free OTTA protocol and still outperforms Tri-Training, further verifying its effectiveness.

\begin{figure}[htpb]
	\centering
	\includegraphics[width=0.7\linewidth,keepaspectratio]{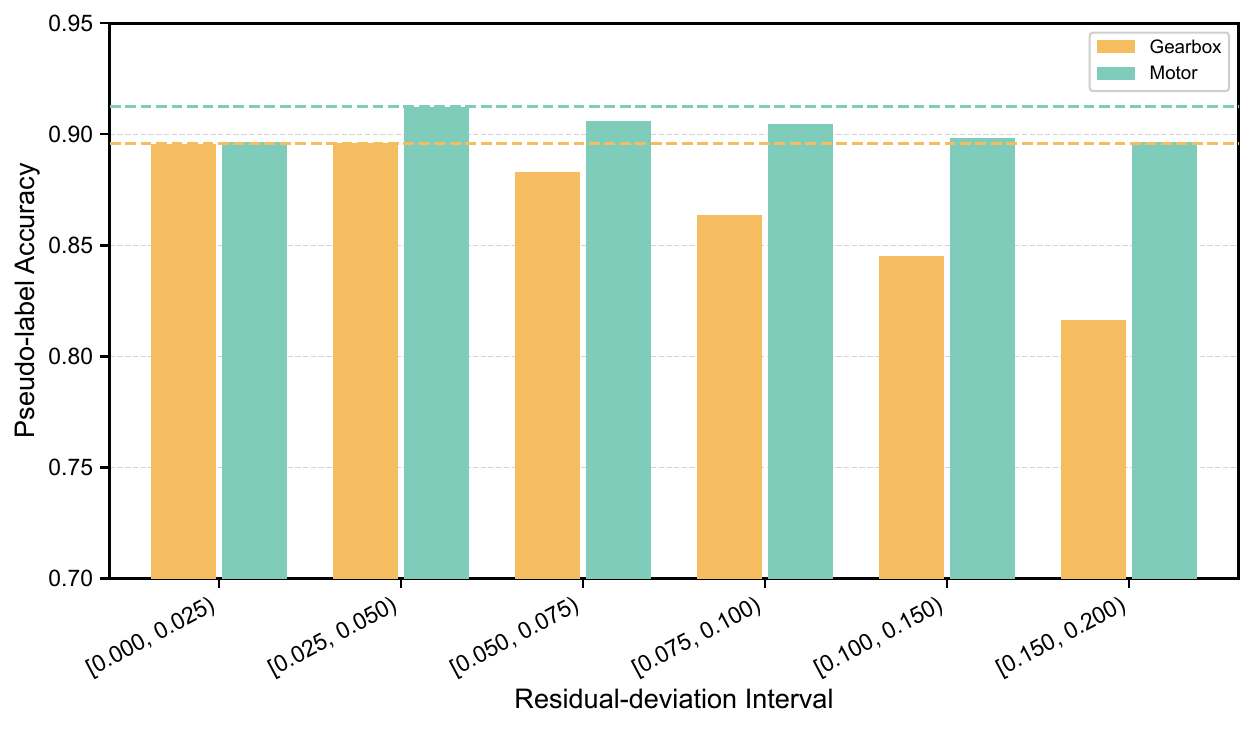}
	\caption{Pseudo-label accuracy within different residual-deviation intervals.}
	\label{residual_pl_acc}
\end{figure}

\subsubsection{Online Mechanism Analysis}
To further investigate the residual condition guidance mechanism in the online stage,
the relationship between residual deviation and sample reliability is quantitatively analyzed by computing pseudo-label accuracy over different residual-deviation intervals.
As shown in Fig.~\ref{residual_pl_acc}, the pseudo-label accuracy on the gearbox dataset exhibits a clear overall decreasing trend as the residual deviation increases, while a similar trend, although less pronounced, is observed on the motor dataset.
These results provide empirical support for using residual deviation as a reliability indicator.

\subsubsection{Transition Stage Analysis}
The FIFO queue may introduce a short-term lag after an operating-condition or fault transition, since its mean response still reflects the preceding state.
Consequently, samples from the new state may be selected more conservatively, which helps reduce unreliable updates and error accumulation during the transition.
As new responses progressively replace the old ones in the queue, the mean adapts to the current state and the selection behavior gradually recovers.

\begin{table}[htpb]
	\centering
	\caption{Efficiency comparison on Gearbox.}
	\label{table_efficiency}

	\begin{threeparttable}
	
	\resizebox{\columnwidth}{!}{%
	\begin{tabular}{l|ccccc}
	\specialrule{0.1em}{3pt}{1pt}
	\specialrule{0.1em}{1pt}{1pt}
	
	\rowcolor{gray!12}
	\textit{Method}
	& \makecell{\textit{Running Time} $\downarrow$\\\textit{(s)}}
	& \makecell{\textit{Throughput} $\uparrow$\\\textit{(samples/s)}}
	& \makecell{\textit{Peak GPU Mem.} $\downarrow$\\\textit{(MB)}}
	& \makecell{\textit{Fwd. Vol.} $\downarrow$\\\textit{(M/batch)}}
	& \makecell{\textit{Bwd. Vol.} $\downarrow$\\\textit{(M/batch)}} \\
	
	\midrule

	\textit{Tent-Full}
	& 1.14 & 21215 & 155 & 6.965 & 6.965 \\
	
	\textit{AdaContrast}
	& 10.85 & 2195 & 265 & 27.861 & 6.965 \\
	
	\textit{CoTTA}
	& 11.76 & 2032 & 391 & 36.154 & 6.965 \\
	
	\textit{DPLOT}
	& 1.48 & 16163 & 298 & 34.827 & 13.931 \\
	
	\textit{SURGEON}
	& 1.61 & 15397 & 242 & 20.896 & 13.931 \\
	
	\textit{OAFD}
	& 1.48 & 16366 & 235 & 13.931 & 6.965 \\

	\textit{Tri-Training}
	& 1.54 & 15588 & 406 & 63.811 & 0.225 \\
	
	\cmidrule(lr){1-6}

	\rowcolor{gray!8}
	\textbf{\MethodName}
	& 1.71 & 14001 & 216 & 13.972 & 6.965 \\
	
	\rowcolor{gray!8}
	\textbf{\MethodName\ (CPU)}
	& 4.79 & 4992 & -- & 13.972 & 6.965 \\
	
	\specialrule{0.1em}{3pt}{1pt}
	\specialrule{0.1em}{1pt}{1pt}
	\end{tabular}%
	}
	
	\begin{tablenotes}
		\footnotesize
		\item[] Notes: $\uparrow$ and $\downarrow$ indicate that higher and lower values are preferred, \\respectively.
	\end{tablenotes}
	\end{threeparttable}
\end{table}

\subsubsection{Computational Efficiency Analysis}
Computational efficiency should also be considered in practical online fault diagnosis.
Table~\ref{table_efficiency} reports both hardware-dependent measurements, including the running time of a complete online stream, throughput, and peak GPU memory usage,
and platform-independent indicators, including the cumulative parameter volumes involved in forward and backward propagation for each online batch.
The proposed \MethodName\ achieves higher diagnostic accuracy while maintaining a favorable balance between diagnostic performance and computational overhead. 
The CPU result further provides a cross-platform reference for practical online deployment.

\color{black}

\subsection{Ablation Study}

To further verify the contribution of each component in the proposed framework, an ablation study is conducted by progressively adding different modules, as shown in Table~\ref{table_ablation}.

\color{black}
\textit{Baseline} denotes the MLP diagnostic model. 
Due to its limited generalization ability, the overall accuracy is only 0.779. 
After introducing the continuous operating-condition adversarial learning strategy with progressive training in the offline stage, 
\textit{CAL} consistently outperforms both \textit{Baseline} and \textit{DANN}, improving the overall accuracy to 0.868.
This result verifies that modeling operating conditions in a continuous manner is more suitable than in a discrete manner for enhancing the generalization ability under time-varying operating conditions.
In the online stage, \textit{CF} updates the student model only with samples of sufficiently high pseudo-label confidence, 
leading to an improvement of 0.035 in overall accuracy over \textit{CAL}.
Furthermore, \textit{CAL+RCG} performs online adaptation by selecting samples according to the residual condition guidance criterion.
It achieves an overall accuracy of 0.910, demonstrating that the residual condition response provides more reliable information for online sample selection than confidence filtering alone. 
Finally, \textit{\MethodName} integrates confidence filtering and residual condition guidance into a complete mechanism, achieving the highest overall accuracy of 0.918. 
Therefore, although \textit{CAL} can improve the offline representation, the online adaptation stage still provides an obvious boost in accuracy, highlighting its effectiveness.
Moreover, introducing the online adaptation mechanisms consistently reduces the ECE compared with \textit{CAL} alone. 
While confidence filtering may encourage sharper predictions and therefore result in a slightly higher ECE for the complete \textit{\MethodName} than for \textit{CAL+RCG}, the complete framework still maintains a relatively low ECE of 4.4\%.

\begin{figure*}[!t]
	\centering
	\sbox{\hyperfigbox}{\includegraphics[width=0.3\textwidth]{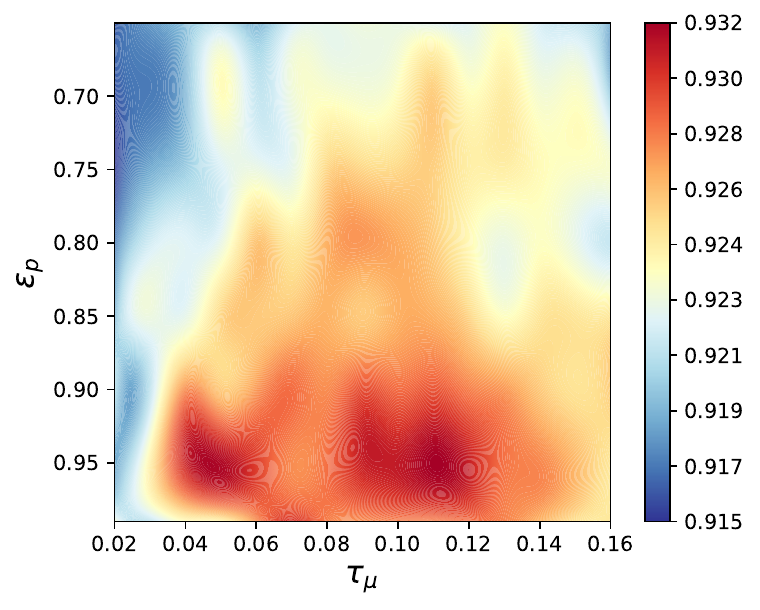}}%
	\subfloat[Effect of $B_{\text{on}}$]{%
		\begin{minipage}[b][\ht\hyperfigbox][c]{0.4\textwidth}
			\centering
			\includegraphics[width=\linewidth]{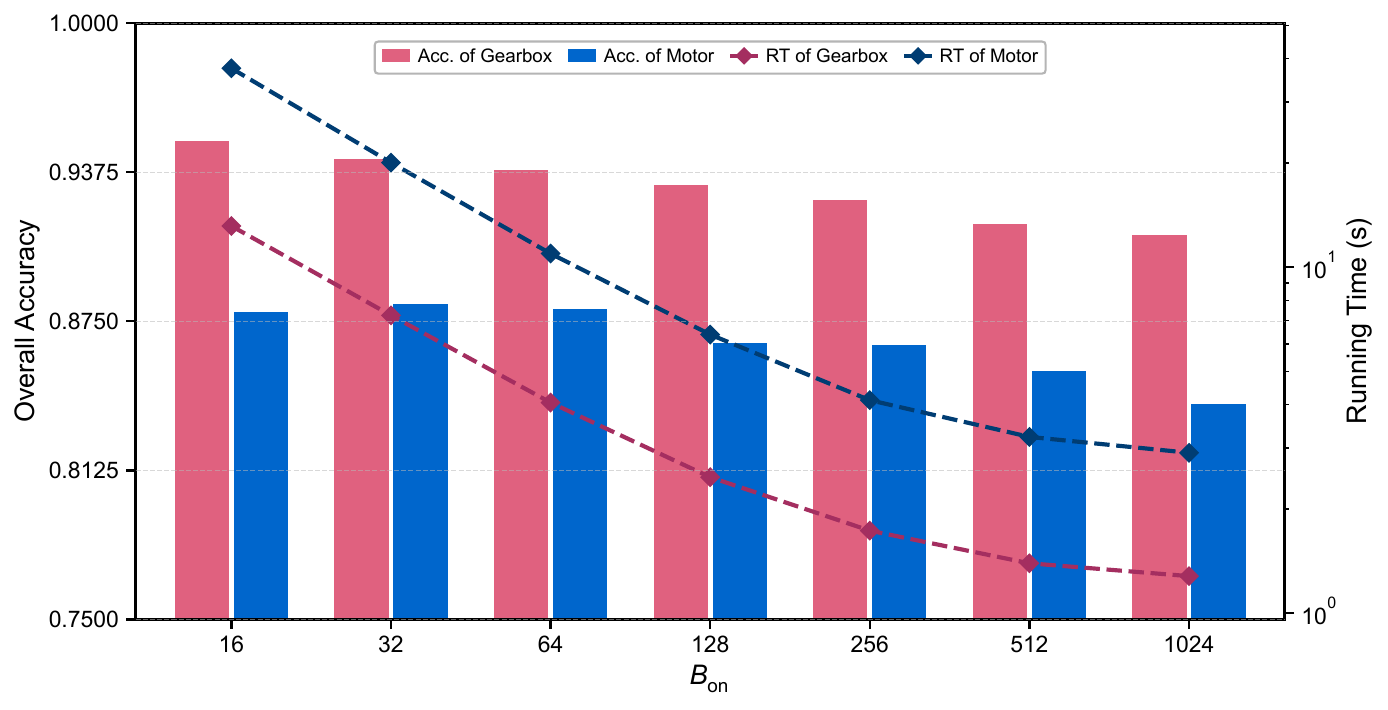}%
		\end{minipage}\label{hyper-1}}
	\hfil
	\subfloat[Effect of $\tau_{\mu}$ and $\epsilon_p$]{%
		\begin{minipage}[b][\ht\hyperfigbox][c]{0.3\textwidth}
			\centering
			\includegraphics[width=\linewidth]{./figures/heatmap_gearbox_thres.pdf}%
		\end{minipage}\label{hyper-2}}
	\hfil
	\subfloat[Effect of $\tau_{\mu}$ and $W$]{%
		\begin{minipage}[b][\ht\hyperfigbox][c]{0.3\textwidth}
			\centering
			\includegraphics[width=\linewidth]{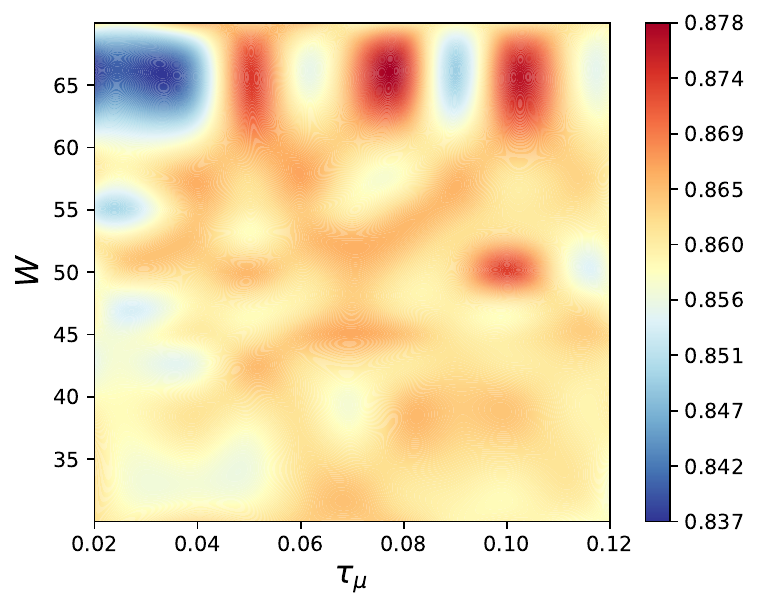}%
		\end{minipage}\label{hyper-3}}
	\caption{Hyperparameter study of the proposed \MethodName.
	The results in (a) are obtained under the gear wear and voltage unbalance fault scenarios, 
	while (b) and (c) are obtained under the gear wear and voltage unbalance fault scenarios, respectively.
	The color in (b) and (c) denotes the diagnostic accuracy.
	}
\end{figure*}

\begin{table}[htpb]
	\centering
	\caption{Effects of components in \MethodName.} 
	\label{table_ablation}
	
	\begin{threeparttable}
	\resizebox{\columnwidth}{!}{%
	\begin{tabular}{l|cc|cc|cc}
		\specialrule{0.1em}{3pt}{1pt}
		\specialrule{0.1em}{1pt}{1pt}
		
		\multirow{2}{*}{\raisebox{-0.6ex}{\textit{Strategy}}} 
		& \multicolumn{2}{c|}{\textit{Offline}} 
		& \multicolumn{2}{c|}{\textit{Online}} 
		& \multicolumn{2}{c}{\textit{Performance}} \\
		\cmidrule(lr){2-7}
		\cmidrule(lr){4-5}
		\cmidrule(lr){6-7}
		
		& \textit{DAL}
		& \textit{CAL}
		& \textit{CF}
		& \textit{RCG}
		& \textit{Acc. (\%)} 
		& \textit{ECE (\%)} \\
		
		\midrule
		
		\textit{Baseline}       
		&  &  &  &  
		& 77.9 
		& 21.0 \\
		
		\textit{DANN}              
		& \cmark &  &  &  
		& 86.2 
		& 10.5 \\

        \cmidrule(lr){1-7}

		\textit{CAL}              
		&  & \cmark &  &  
		& 86.8$_{(+0.0)}$ 
		& 9.5$_{(+0.0)}$ \\
		
		\textit{CAL+CF}            
		&  & \cmark & \cmark & 
		& 90.3$_{(+3.5)}$
		& 4.4$_{(-5.1)}$ \\
		
		\textit{CAL+RCG}            
		&  & \cmark &  & \cmark 
		& 91.0$_{(+4.2)}$
		& \textbf{3.7}$_{(-5.8)}$ \\
		
		\rowcolor{gray!10}
		\textbf{\MethodName\textsuperscript{*}}     
		&  & \cmark & \cmark & \cmark 
		& \textbf{91.8}$_{(+5.0)}$
		& 4.4$_{(-5.1)}$ \\
		
		\specialrule{0.1em}{3pt}{1pt}
		\specialrule{0.1em}{1pt}{1pt}
	\end{tabular}
	}
	
	\begin{tablenotes}
		\footnotesize
		\item[] Notes: The results are obtained on the MCC5-THU Gearbox Dataset. 
		\textbf{DAL} denotes discrete operating-condition adversarial learning implemented by DANN,
		\textbf{CAL} denotes continuous operating-condition adversarial learning with progressive training, 
		\textbf{CF} denotes confidence filtering, and 
		\textbf{RCG} denotes residual condition guidance for online sample selection.
	\end{tablenotes}
	\end{threeparttable}
\end{table}

In conclusion, the above results indicate that each component of the proposed framework makes a contribution the overall performance.
Combining the two online selection criteria improves the overall diagnostic accuracy and the reliability of model updates.

\subsection{Hyperparameter Study}

To investigate the robustness of the proposed CGSA, hyperparameter experiments are conducted under the two representative faults, gear wear and voltage unbalance,
using the same settings as in the comparison experiments.

As illustrated in Fig.~\ref{hyper-1}, small and moderate batch sizes generally provide higher diagnostic accuracy because more frequent updates enable the model to respond more promptly to distribution changes.
However, they also substantially increase the running time because they require more optimization steps. The resulting computational cost may limit continuous online processing.
Fig.~\ref{hyper-2} shows the joint sensitivity to the residual condition consistency threshold $\tau_{\mu}$ and the pseudo-label confidence threshold $\epsilon_p$. 
An overly loose threshold combination may introduce unreliable pseudo-labels, whereas excessively strict thresholds reduce the number of samples available for online adaptation. 
Similarly, Fig.~\ref{hyper-3} shows the joint influence of $\tau_{\mu}$ and the queue length $W$ within a certain range, 
where a small $W$ provides insufficient historical information, leading to a slight decrease in performance.

Overall, the proposed \MethodName{} maintains relatively robust diagnostic performance over reasonable ranges of the evaluated hyperparameters, supporting its applicability.
\color{black}

\section{Conclusion}
In this paper, a {\MethodName} framework has been proposed to tackle the problem of fault diagnosis of dynamic systems under unknown operating conditions.
In the offline stage, a continuous operating-condition adversarial learning strategy with progressive training has been developed to enhance cross-condition generalization.
During online deployment,
residual condition responses and pseudo-labels provided by the fixed source pre-trained model have been exploited to identify reliable streaming samples for student model updates.
The proposed mechanism has reduced error accumulation and improved the stability of online adaptation without requiring access to online ground-truth labels.
Extensive experiments on realistic industrial gearbox and motor platforms have demonstrated that 
the proposed method achieves a favorable balance between diagnostic accuracy and test-time efficiency.

Although the proposed method has demonstrated promising performance under unknown operating conditions, 
its capability to identify previously unknown fault categories requires further investigation.
Future work will therefore focus on extending the proposed framework to unknown fault detection and 
evaluating its robustness in more realistic industrial scenarios and more diverse operating environments.
\color{black}

\bibliographystyle{IEEEtran}
\bibliography{mybibfile}

\end{document}